\documentclass[preprint2,times,tighten]{aastex63}
\usepackage{amsmath,amstext,amssymb}
\usepackage{wasysym} 						
\usepackage{verbatim}
\usepackage{soul}
\usepackage[T1]{fontenc}
\usepackage{apjfonts} 
\usepackage[figure,figure*]{hypcap}
\usepackage{rotating}
\usepackage{mathtools}
\usepackage{multirow}
\shorttitle{Modeling Turbulent Material in the CGM}
\shortauthors{Buie et al.}

\begin{document}

\title{Modeling Photoionized Turbulent Material in the Circumgalactic Medium III: Effects of Co-rotation and Magnetic Fields}

\author{Edward Buie II}
\affiliation{Arizona State University School of Earth and Space Exploration, P.O. Box 871404, Tempe, AZ 
85287, USA}
\author{Evan Scannapieco} 
\affiliation{Arizona State University School of Earth and Space Exploration, P.O. Box 871404, Tempe, AZ 
85287, USA}
\author{G. Mark Voit} 
\affiliation{Michigan State University Department of Physics and Astronomy, East Lansing, MI 
48824, USA}

\begin{abstract}

Absorption-line measurements of the circumgalactic medium (CGM) display a highly non-uniform distribution of lower ionization state species accompanied by more widespread higher ionization state material. This suggests that the CGM is a dynamic, multiphase medium, such as arises in the presence of turbulence.  To better understand this evolution, we perform hydrodynamic and magneto-hydrodynamic (MHD) simulations of the CGM surrounding Milky Way-like galaxies. In both cases, the CGM is initially in hydrostatic balance in a $10^{12}$~M$_{\odot}$ dark matter gravitational potential, and the simulations include rotation in the inner halo and turbulence that decreases radially. They also track ionizations, recombinations, and species-by-species radiative cooling in the presence of the redshift-zero UV background, employing the MAIHEM non-equilibrium chemistry package. We find that after 9~Gyrs of evolution, the presence of a magnetic field leads to an overall hotter CGM, with cool gas in the center where magnetic pressure dominates. While the non-MHD run produces more cold clouds overall, we find similar \ion{Si}{4}/\ion{O}{6} and \ion{N}{5}/\ion{O}{6} ratios between the MHD and non-MHD runs, which are both very different from their equilibrium values. The non-MHD halo develops cool, low angular momentum filaments above the central disk, in comparison to the MHD run that has more efficient angular momentum transport, especially for the cold gas which forms a more ordered and extended disk late into its evolution.

\end{abstract}


\section{Introduction} \label{introduction}

The circumgalactic medium (CGM) is the predominantly ionized atmosphere of diffuse baryons that extends hundreds of kpc from a galaxy, contained by the overall dark matter gravitational potential. It is responsible for regulating the galaxy's evolution through processes such as galactic accretion, starburst-driven winds, and feedback from active galactic nuclei (AGN) \citep[e.g.][]{2013lilly,2015voit,2015crighton,fox2017gas,2017muratov,2017ARA&A..55..389T}. While this medium is often modeled as a purely hydrodynamic fluid in numerical simulations, due to its ionized nature, its dynamics and chemistry can be strongly affected by the presence of magnetic fields.

Observations have shown that magnetic fields exist in the CGM at dynamically important field strengths \citep{1982ApJ...263..518K,1984ApJ...279...19W}. By correlating Faraday rotation measurements of high-redshift radio sources with those of foreground galaxies with observed \ion{Mg}{2} absorption, the line-of-sight component of the $B$-field in the CGM has been found to be a few $\mu$G \citep[e.g.][]{2008Natur.454..302B,2008kronberg,2010ApJ...711..380B,2012ApJ...761..144B,2014ApJfarnes,2020malik,2020MNRAS.496.3142L}. 

In addition to this, absorption lines have shown the CGM to be multiphase: possessing cold, $T \approx 10^{4}$~K, gas traced by ions such as \ion{H}{1} and \ion{Mg}{2} in rough pressure equilibrium with a hotter, $T \approx 10^{6}$~K phase that is traced by \ion{O}{6} and \ion{Ne}{8} \citep[e.g.][]{Chen1998,1998steidel,rudie2012gaseous,tuml2013ApJ...777...59T,heckman2017cos}. Measurements of ion ratios, such as \ion{Si}{4}/\ion{O}{6} and \ion{N}{5}/\ion{O}{6} have also been able to probe the $T \approx 10^{5}$~K transitioning, intermediate phase gas in the CGM \citep{werk2016ApJ...833...54W}, yielding results that are inconsistent with models that assume that the various phases are in chemical equilibrium.

However, such observations have been difficult to interpret in detail, largely owing to the low densities in the CGM, which lead to low detection rates \citep{2017ARA&A..55..389T}. This has left open questions about the magnetic field structure and its evolution  \citep{1982ApJ...263..518K,1984ApJ...279...19W}, as well as its influence on CGM dynamics, leading to several numerical simulations focused on the impact of magnetic fields on the gas beyond galaxies.

On somewhat larger scales, \citet{2009MNRAS.399..497D} ran magneto-hydrodynamic (MHD) cosmological simulations of galaxy cluster environments and found that the dissipation of the magnetic field could play an important role in the determination of the predicted magnetic field profile. Years later, \citet{2012MNRAS.422.2152B} ran cosmological MHD simulations to understand the evolution of $B$-fields in the context of Milky Way-like galactic halo formation. By injecting a magnetic field at a low rate of 1~nG~Gyr$^{-1}$, they found it exponentially amplified to $\mu$G level near the halo center due to shocks and turbulent dynamo action, the process in which the magnetic field strengths are amplified by the stretching and twisting of magnetic field lines from turbulence \citep{1968JETP...26.1031K,1980opp..bookR....K,1990alch.book.....Z,Brandenburg_2012}. This rapid amplification was also studied in galaxy clusters in \citet{2014MNRAS.445.3706V}.

Other cosmological simulations of Milky Way-like galaxies have similarly found initial $B$-fields to be quickly amplified to $\mu$G strengths via turbulent dynamo action. This is initially triggered by feedback processes (e.g. star formation, active galactic nuclei, supernovae, etc.), and maintained by large differential rotation of the central galaxy \citep[e.g.][]{2014ApJ...783L..20P,2016MNRAS.457.1722R,2017MNRAS.469.3185P,2017MNRAS.471.2674R,2017MNRAS.472.4368R}. However, these simulations have been predominantly focused on the central galaxy and the immediate region around it, not the entire CGM. 

More recently, \citet{2018MNRAS.475..624N} ran large-scale MHD simulations as part of the IllustrisTNG project \citep{2018MNRAS.473.4077P} and found a correlation between the ratio of magnetic pressure to thermal pressure in the halo gas and the (g-r) color of the central galaxy. Also as part of the IllustrisTNG project, \citet{2018MNRAS.480.5113M} investigated magnetic properties of galaxy clusters and galactic halos, finding the ratio of magnetic pressure to thermal pressure reaches a maximum of 3 in the center and declines with increasing distance. Moreover, \citet{Pakmor_2020} used cosmological zoom-in simulations of Milky Way-like galaxies to understand the $B$-field evolution in the context of the CGM. They found that outflows initially magnetize the CGM, and that these fields were further amplified via turbulent dynamo action to a strength of 0.1~$\mu$G at the virial radius. \citet{2020MNRAS.498.2391N} also conducted cosmological MHD simulations using the IllustrisTNG model to investigate the cold gas in the CGM of massive halos with $M \gtrsim 10^{13}$~M$_{\odot}$, showing that such cold clouds possessed magnetic pressures that were at least 10 times greater than their thermal pressures.

In this work, we carry out the first simulation of the evolution of a magnetized Milky Way-like galactic halo that includes full non-equillibrium chemistry, rotation, and turbulence. We complement this simulation with a hydrodynamic case that is otherwise identical, allowing us conduct direct comparisons in a controlled way to better understand the impact of magnetic fields. Together, these simulations allow us to make detailed models of both the multiphase nature of the magnetized CGM, and its observable properties shown by absorption line tracers that probe a wide variety of ionization states.

The structure of this work is as follows. In Section~\ref{methods} we discuss the model used to simulate this environment. In Section~\ref{results} we compare between the co-rotating hydrodynamic case and the co-rotating MHD case, and look at how these considerations influence the phase, magnetic field structure/strengths, kinematics of ions, as well as the non-equilibrium chemistry that develops in the CGM. We conclude this work by summarizing and discussing the results in Section~\ref{summary}.

\section{Methods} \label{methods}

\subsection{The Modified MAIHEM Code} \label{maihem}
To carry out non-equillibrium MHD simulations of  the magnetized CGM, we used Models of Agitated and Illuminated Hindering and Emitting Media (MAIHEM\footnote{http://maihem.asu.edu/}). MAIHEM is a cooling and chemistry package built using FLASH (Version 4.5), an open-source hydrodynamics code \citep{fryxell2000flash}. This package models the hydrodynamics in three-dimensions (3D) and evolves a non-equilibrium chemistry network of 65 ions, including hydrogen (\ion{H}{1} and \ion{H}{2}), helium (\ion{He}{1}--\ion{He}{3}), carbon (\ion{C}{1}--\ion{C}{6}), nitrogen (\ion{N}{1}--\ion{N}{7}), oxygen (\ion{O}{1}--\ion{O}{8}), neon (\ion{Ne}{1}--\ion{Ne}{10}), sodium (\ion{Na}{1}--\ion{Na}{3}), magnesium (\ion{Mg}{1}--\ion{Mg}{4}), silicon (\ion{Si}{1}--\ion{Si}{6}), sulfur (\ion{S}{1}--\ion{S}{5}), calcium (\ion{Ca}{1}--\ion{Ca}{5}), iron (\ion{Fe}{1}--\ion{Fe}{5}), and electrons. This includes solving for dielectric and radiative recombinations, collisional ionizations with electrons, charge transfer reactions, and photoionizations by a UV background. 
%\textbf{We plan to make MAIHEM publicly available in the coming months.}

MAIHEM was originally developed in \citet{gray2015atomic}, later improved upon with the inclusion of an ionizing background in \citet{gray2016atomic}, and further updated with several charge transfer reactions, radiative recombination rates, and dielectronic recombination rates from \citet{aldrovandi1973radiative,shull1982ionization,arnaud1985updated} in \citet{gray2017effect}. Physical cooling processes are included down to 5000~K.

Most recently, \citet[][hereafter B20]{Buie_II_2020} modified the MAIHEM code to simulate this gas in a dark-matter \citet[][hereafter NFW]{Navarro1996} gravitational potential, given by:
\begin{equation}
\rho_{\rm NFW}(r) = \frac{\rho_0}{\frac{r}{R_s} 
\left(1 + \frac{r}{R_s} \right)^2},
\end{equation}
where $\rho_0  = M_{\rm halo} \left[4\pi R_s^{3} \left[ \ln(1+c) - \right. \right.$ $\left. \left. c/(1+c) \right] \right]^{-1}$ is the dark matter density normalization, $M_{\rm halo}$ is the mass of the halo, $R_s = R_{\rm vir}/c$ is the scale radius while $R_{\rm vir}$ is the virial radius of the halo, and $c$ is the concentration parameter of the halo. The turbulence, which is used as our driving feedback mechanism, is artificially driven according to the following equation, 
\begin{equation}
a_{x,y,z} = a^0_{x,y,z} \left(\frac{r + 0.3 R_{\rm vir}}{0.5 R_{\rm vir}}\right)^{\alpha},
\label{equ:trend}
\end{equation} 
where $a^0_{x,y,z}$ is the original acceleration term resulting from the direct Fourier transform of the solenoidal modes ($\nabla \cdot F = 0$, where $F$ is the artificial forcing term used in the momentum equation) \citep{pan2010}, $r$ is the radius to a cell, and $\alpha$ is a dimensionless parameter that controls how the acceleration behaves with radius. These solenoidal modes depend on the driving scale of turbulence, total injected energy, and the autocorrelation time for the Ornstein-Uhlenbeck (OU) process that these are modeled as \citep{uhlenbeck1930theory}. 

\citet{gray2016atomic} details the equations solved in MAIHEM which are invariant under the transformation $x \rightarrow \lambda x,\ t \rightarrow \lambda t,\ \rho \rightarrow \rho/\lambda$ meaning the final steady-state abundances depend only on the product $nL$ of the mean density and the driving scale of turbulence, the one-dimensional (1D) velocity dispersion of the gas, $\sigma_{\rm 1D}$, and the ionizing extragalactic UV background (EUVB). 

We utilize the unsplit staggered mesh (USM) algorithm which solves multidimensional ideal and non-ideal MHD problems on a Cartesian grid and the equations for these calculations are available in \citet{2013leeJCoPh.243..269L}. This study specifically uses the ideal MHD solvers, and we also make use of a hybrid Riemann solver that uses the Harten Lax and van Leer (HLL) solver \citep{einfeldt1991godunov} in places with strong shocks or rarefactions and the Harten--Lax--van Leer--Contact (HLLC) solver \citep{toro1994restoration,tororiemann} in smoother flows to stabilize the code as turbulence ensues. We refer the reader to \citet{gray2015atomic} and \citet{gray2016atomic} for further details. 

\subsection{Model Parameters} \label{model_params}

We conducted a suite of numerical simulations that model an 800~kpc box with periodic boundaries using Static Mesh Refinement (SMR) to accurately capture important structures that develop in the halo as turbulence ensues. The simulations are run in a Milky Way mass dark matter halo of 10$^{12}$~M$_{\odot}$, with a virial radius of 220~kpc and a concentration parameter of 10. Note that the virial radius remains constant with time, and the simulations are not run in a comoving domain.

Cosmological parameters are solely used to simulate the gravitational acceleration of the baryonic matter and are from the Planck 2018 Collaboration \citep{planck2018}. They are $h$ = 0.674, $\Omega_m$ = 0.315, $\Omega_b$ = 0.049, and $\Omega_{\Lambda}$ = 0.685, where $h$ is the Hubble constant in units of 100 km s$^{-1}$ Mpc$^{-1}$, and $\Omega_m$, $\Omega_b$, and $\Omega_{\Lambda}$, are the total matter, baryonic, and vacuum densities, respectively, in units of the critical density. 

The resolution was set to that in B20, meaning that for $R \gtrsim$ 300~kpc the domain was at a resolution of 64$^{3}$ which translates to 12.5~kpc, for 300 $\gtrsim R \gtrsim$ 250~kpc the resolution was 128$^3$ which translates to 6.2~kpc, 250 $\gtrsim R \gtrsim$ 225~kpc the resolution was 256$^3$ which translates to 3.1~kpc, and for 225~kpc $\gtrsim R$ the resolution was 512$^{3}$ which translates to 1.6~kpc. 

Note that \citet[][hereafter B18]{Buie_II_2018} used a smaller, sub-parsec resolution and a smaller turbulent driving scale as compared to the runs presented here. Both sets of simulations inject turbulence with stirring modes between $L/3 \leqslant 2\pi/k \leqslant L$, but in B18, $L$ was set to be on the scale of the box, 100~pc, while here it is chosen to be 30~kpc. This is about the size of the Milky Way disk as we assume such turbulence would be primarily driven by outflow/inflow processes on the scale of the disk. However, it is misleading to match these physical scales directly, because for a fixed ionization parameter, the results of the isotropic turbulence simulations such as presented in B18, depend on the product of the density and the driving scale $nL$, rather than directly on $L$. As the densities that were used in this previous work were orders of magnitude larger than those found in the study described here, $nL$ is comparable between the two sets of simulations, even though $L$ is not.

Likewise, it is the resolution in units of $L$ rather than in physical units that is the key measure of resolution. In B18 the resolution was $\Delta x=L/128$ while in the current case it is slightly lower, $\Delta x \approx L/20$, where 20 results from the resolution divided by the driving scale of turbulence. However, B20 also conducted a resolution study by doubling the resolution to 1024$^{3}$ or 0.8~kpc and found that while the cool structures that developed were smaller in scale, the overall properties and conditions in the halo such as temperatures, densities, and resulting column densities, remained in agreement with the original resolution counterpart. MHD simulations as large as 1024$^3$ that are run for 9 Gyrs are prohibitively expensive, and we predict that in such simulations, the resulting cool structures would be smaller but the overall properties would be largely similar to the runs presented here.
 
We initialize the medium with a Keplerian circular velocity modified by the observational findings from the \ion{Mg}{2} studies of \citet{2017_Ho}. Specifically, we use their Equation A2 to inform how the velocities should fall off along the minor axis, which is chosen to be the $y$ axis in our runs. This gives a rotational velocity of 
\begin{equation} 
    v_{\rm rot} = f_{\rm rot} \sqrt{\frac{4 \pi G r^2 \rho_{\rm NFW}(r)}{3}}  H_s(y),
    \label{equ:rotation}
\end{equation}
%\begin{equation}
%    v_z = \frac{-x}{r_{2D}} \sqrt{\frac{4 \pi G r^2 \rho_{NFW}}{3}} H_s,
%\end{equation}
where $G$ is the gravitational constant, and $H_s \equiv \exp{(-\lvert y \rvert/{\rm 50~kpc})}$ is the dependence of the circular velocity with scale height, and $f_{\rm rot}$ is an overall scaling parameter we set to 60\%. In this case the spin parameter of the gas, $\lambda$ defined as 
\begin{equation}
    \lambda = \frac{J}{\sqrt{2} M_{\rm gas} V R},
\end{equation}
where $J$ is the total angular momentum within a sphere of virial radius $R$ along with its gas mass $M_{\rm gas}$. $V$ is the circular velocity at the virial radius, $V = \sqrt{GM_{\rm total}/R}$. The chosen $f_{\rm rot}$ results in $\lambda \approx 0.03$ in our halos, in line with numerical studies of galactic dark matter halos \citep{2001ApJ...555..240B,2008MNRAS.391.1940M,2013MNRAS.429.3316B}.

In all our runs, the gas was initialized with a fractional ion abundance of 0.3~Z$_{\odot}$ in collisional ionization equilibrium (CIE) at the virial temperature of the halo, $T = 1.2 \times 10^{6}$~K and average sound speed of 166~km~s$^{-1}$. Following initialization, the medium is allowed to evolve according to the non-equilibrium chemistry. We note that our simulations assume the metallicity to be constant throughout the halo, while observations support a CGM that is significantly chemically inhomogeneous \citep{Zahedy2019,buie2020interpreting}. We also choose a redshift zero \citet[][hereafter HM2012]{2012ApJ...746..125H} EUVB whose specific intensity was normalized to $8.23 \times 10^{-24}$ erg cm$^{-2}$ s$^{-1}$ Hz$^{-1}$ sr$^{-1}$ at the Lyman limit to irradiate the gas in the runs.

\begin{figure*}
    \includegraphics[width=1.0\linewidth,height=0.9\textheight]{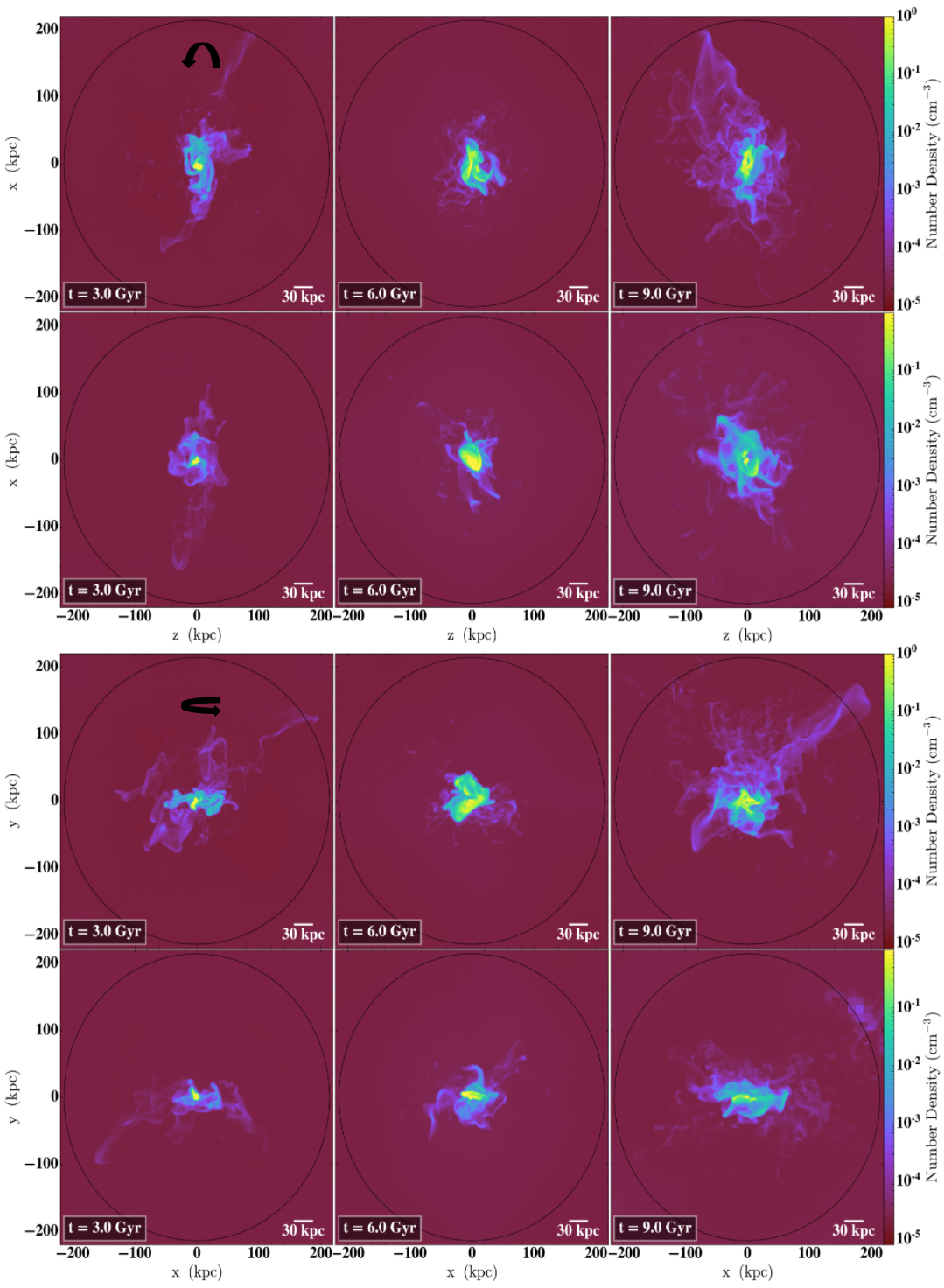}
    \caption{Projections of mass-weighted number densities at 3~Gyrs (left), 6~Gyrs (middle), and 9~Gyrs (right). The first and second rows show these from a face-on perspective for the Hydro and MHD Rot halos, respectively. The third and forth rows show these from an edge-on perspective for the the Hydro and MHD Rot halos, respectively. We show black arrows to indicate the direction of rotation in the halo and black circles to show the virial radius at $R_{\rm vir} \approx 220$ kpc.}
    \label{fig:ndens_slices}
\end{figure*}

We conduct runs that build off of the previous High run from B20, which was able to match many ion observations of nearby star-forming galaxies' CGMs. That run was conducted by setting $\alpha$ in Equation \ref{equ:trend} to -1, and used the same driving scale for turbulence, 30~kpc. This ensured turbulent stirring was strongest towards the center and fell off with radius. Observationally, non-thermal motions that are derived from Voigt profiles are typically smaller for low-ionization state ions, which are more often found near the host, than in higher ionization state ions which may reside at a range of radii from the central galaxy \citep{tuml2013ApJ...777...59T,churchill2015direct,werk2016ApJ...833...54W,2017ApJ...835...52F}. However this may be largely due to the fact low-ionization-state ions occur preferentially in colder material with higher physical densities than higher-ionization-state ions, meaning that at the same column depth they will probe smaller physical scales, wherein turbulence is much smaller. Additionally, \citet{buie2020interpreting} conducted a MCMC investigation of absorbers in the CGM of nearby galaxies and found the turbulent velocities had no definitive radial trend, but instead covered a range in $\sigma_{\rm 1D}$ between $11 - 60$~km~s$^{-1}$, consistent with our turbulent stirring.

The turbulent stirring employed here, and in the previous B20 study, also allows us to more accurately capture the stronger central feedback of actively star-forming galaxies, as well as the impact of gas accreted from the intergalactic medium, which is likely to deposit the most energy {\em per unit volume} near the center of the halo. This choice in turbulent stirring led to an average one-dimensional velocity dispersion $\sigma_{\rm 1D} = 41$~km~s$^{-1}$. In addition to this turbulent stirring treatment, the aforementioned co-rotation prescription was also considered in the run which we call the Hydro Rot run. 

The second run we conducted further included a seed magnetic field in the z-direction of 0.1$~\mu$G run in ideal MHD. This choice of initial magnetic field strength translated to a plasma $\beta$, defined as the ratio of thermal pressure to magnetic pressure, of $\approx$ 170. It also used the same turbulent stirring and co-rotation prescriptions and is called the MHD Rot run.

Previously in B20, the simulations were run for 3~Gyr until they reached a global equilibrium such that the change in total energy was less than 0.25\% per timestep. With the addition of co-rotation, we ran them for 9~Gyrs as interesting features developed post 3~Gyrs and we wanted to capture the full saturation period of the magnetic field in the MHD run. 

Note that these simulations are run in a static domain for 9~Gyrs which do not emulate the mass growth that would likely be experienced by a real galaxy in this evolutionary time. Such mass growth would likely lead to increased turbulent motions as new gas enters the CGM. Furthermore, these interactions would likely increase the magnetic field in these areas due to the twisting of magnetic field lines. 

\begin{figure*}
    \includegraphics[width=1.0\linewidth]{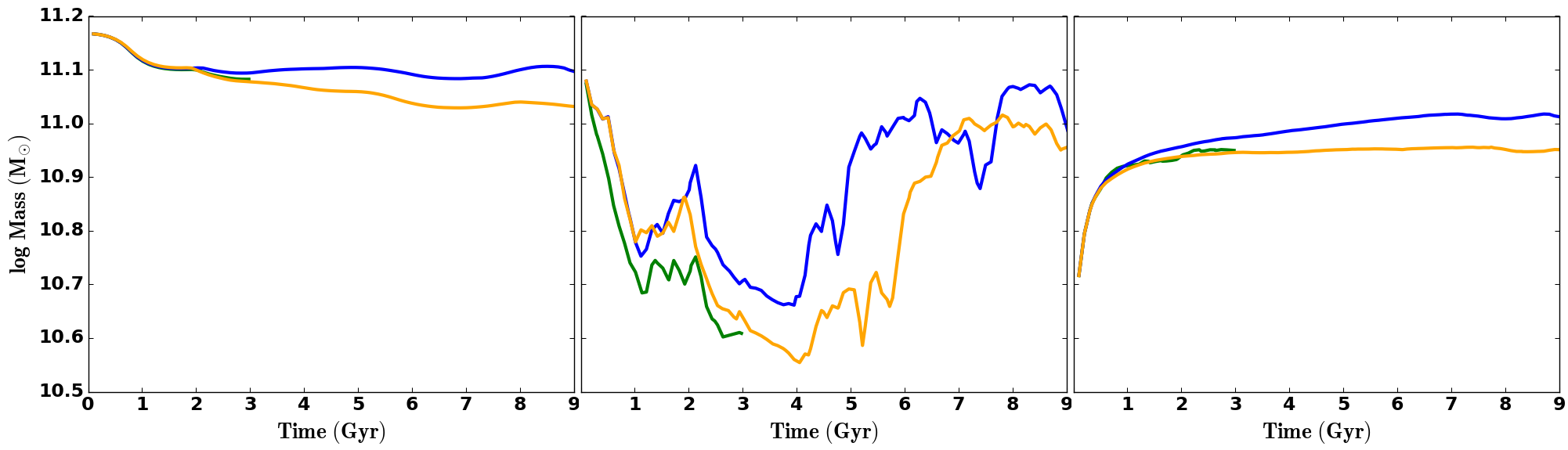}
\caption{Panels show the gas mass vs. time for the Hydro Rot (blue) and MHD Rot (gold) runs, as compared to our run without rotation from B20, the High run (green). The left panel shows the total gas mass within $R_{\rm vir}$. The middle panel shows the gas mass between 12~kpc and $R_{\rm vir}$, omitting the central region. The right panel shows the gas mass within $R_{\rm vir}$ with $T$~<~10$^{5}$~K and includes the central region.}
    \label{fig:mass_combine}
\end{figure*}

\section{Results} \label{results}

\subsection{Overall Evolution} 

\begin{figure}
    \includegraphics[width=1.0\linewidth]{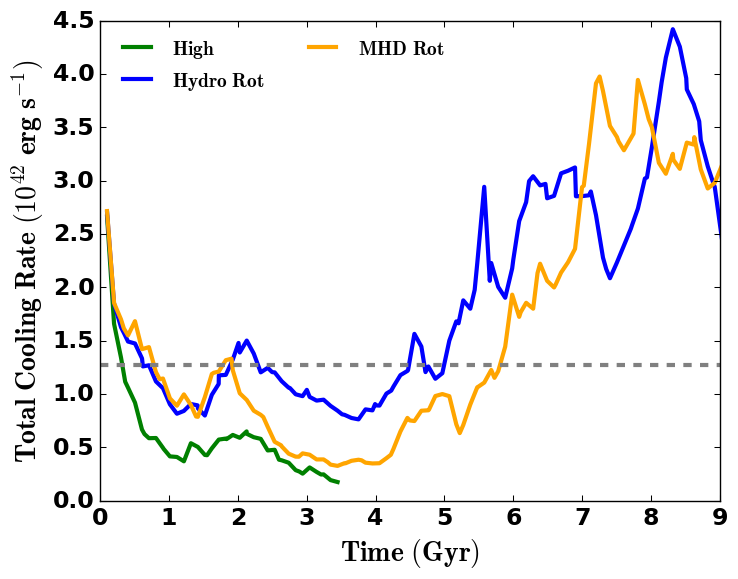}
    \caption{Total cooling in the CGM halo vs. time for the Hydro Rot (blue) and MHD Rot (gold) runs, as compared to our run without rotation from B20, the High run (green). We omit data within 12~kpc as this material primarily traces gas within the galaxy. The grey dashed line shows the turbulent energy injection rate from B20 that was found to be 1.3~$\times$~10$^{42}$~erg~s$^{-1}$ for the High run.}
    \label{fig:cool_mhd_rot_evo}
\end{figure}

To give the reader visual context for the changes in structure as our halos evolve, we show projections of the mass-weighted number density from face-on and edge-on perspectives for the Hydro Rot and MHD Rot halos in Figure~\ref{fig:ndens_slices}, at 3, 6 and 9~Gyrs. Furthermore, Figure~\ref{fig:mass_combine} shows the evolution of the gas mass within $R_{\rm vir}$ in our new simulations, compared to the High simulation from B20, the mass between 12~kpc and $R_{\rm vir}$, as well as the mass in gas with $T < 10^{5}$~K. As in our previous simulations, gas within $r \approx$~30~kpc cools in the initial $\approx$~500~Myrs, forming a low-pressure region in the center of the halo, which promotes an accretion flow of gas to the inner 12~kpc which may be seen from the middle panel of Figure~\ref{fig:mass_combine}. Such cooling may also be inferred from the rapid rise in cool gas observed in the right panel of this figure. Gas near the virial radius is pushed beyond it by the initial turbulence, a feature indicated by a decrease in mass shown in the left panel of this figure.

\begin{figure}
    \includegraphics[width=1.0\linewidth]{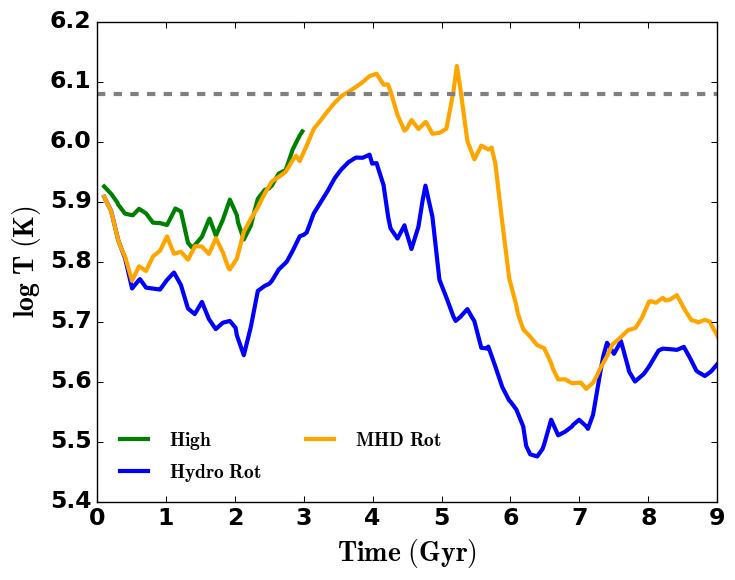}
    \caption{Mass-weighted average temperature in the virial halo vs. time for the Hydro Rot (blue) and MHD Rot (gold) runs, as compared to our run without rotation from B20, the High run (green). We omit data within 12~kpc as this material primarily traces gas within the galaxy.}
    \label{fig:temp_mass_weight_evo}
\end{figure}

The accretion flow results in an accretion shock that forms near the core and gradually moves outward, dispersing the gas near the core throughout the halo. One observes this feature by looking at the middle panel of Figure~\ref{fig:mass_combine} that shows a rise in gas mass between 1 and 2~Gyrs. We also see the initialization of the turbulence, followed by the accretion shock, both push about 14\% of the baryonic mass beyond $R_{\rm vir}$ in the first 2~Gyrs. 

\begin{figure*}
    \includegraphics[width=1.0\linewidth]{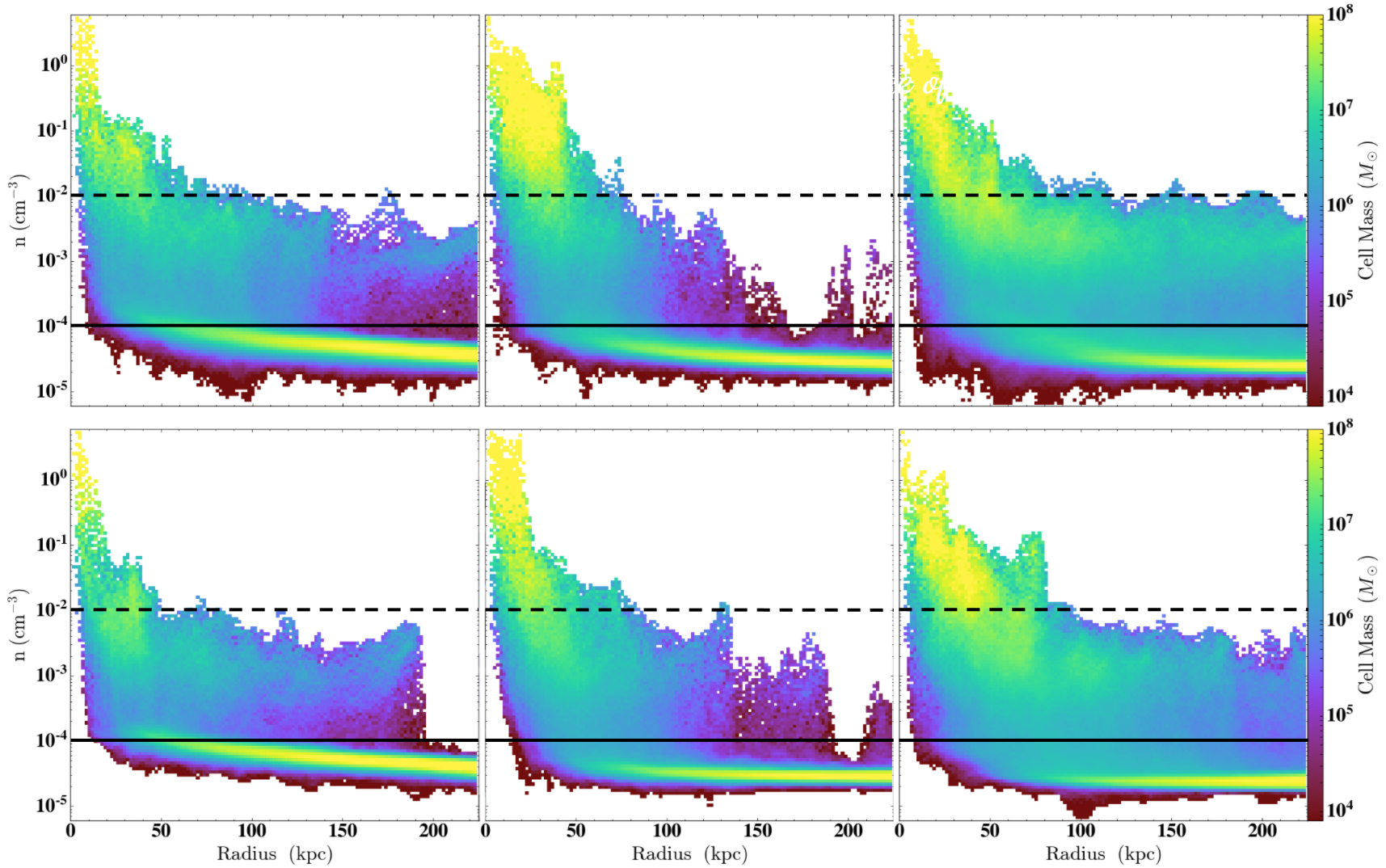}
    \caption{Number density vs. radius showing the total mass in a bin for the Hydro Rot (top) and MHD Rot (bottom) runs. We specifically show these at 3 (left), 6 (middle), and 9 (right) Gyrs. Plots are made using 128 bins for the x and y quantities. For reference, we include a black dashed line to show the number density at 10$^{-2}$~cm$^{-3}$ and a solid line to show this at 10$^{-4}$~cm$^{-3}$.}
    \label{fig:dens_mhd_rot}
\end{figure*}

At approximately 2~Gyrs, the accretion shock leaves the halo. Following this, between 2 - 4~Gyrs, both of the new runs follow similar evolutionary trends in their mass confined to the intermediate region between 12~kpc and $R_{\rm vir}$ with the MHD case showing lower mass during this time. This results primarily from the mass loss that occurs as it leaves the halo, which may be seen from the downward slope in the gold line of the left panel. This may in part be influenced by additional pressure support from the amplified magnetic field, described in more detail in $\S3.3$. The amount of gas cooler than 10$^{5}$ also steadily rises in both runs for the remainder of the evolution.

As ambient gas in the MHD halo is at lower densities, it is unable to cool as efficiently as compared to the Hydro Rot run. This is shown by the total cooling in Figure~\ref{fig:cool_mhd_rot_evo} in additon to the lower amount of cold mass found in this run which may be seen in the right panel of Figure~\ref{fig:mass_combine}. Mass loss is prolonged in the MHD Rot run, continuing on for nearly 7~Gyrs. The total gas mass fluctuates at $\lesssim$~1.1~$\times$~10$^{11}$~M$_{\odot}$ in the MHD Rot run, compared to the Hydro Rot halo which instead fluctuates between 1.2 - 1.3~$\times$~10$^{11}$~M$_{\odot}$ for the remainder of its evolution.

We find that the initial co-rotation along with the continuously driven turbulence adds heat to the hot component, thereby raising its pressure and entropy. We see these as subsonic motions on the order of $\approx$~10~km~s$^{-1}$ and they persist throughout the evolution of both runs. This increased pressure and entropy causes hot gas to buoyantly rise beyond the virial radius. Some of this gas travels far enough to re-enter the 800 kpc box from opposite sides which starts at $\approx$~4~Gyrs and is a consequence of our periodic boundary conditions. We consider this to be a numerical artifact and it does not influence the results of this study. The MHD Rot run possesses a hotter ambient medium with higher levels of entropy and pressure, thus more mass leaves its halo. This rising pressure and entropy are discussed in more detail in the following section.

When looking at the mass-weighted average temperature, shown in Figure~\ref{fig:temp_mass_weight_evo}, we see a positive slope that approaches $T_{\rm vir}$ in the MHD case and $< T_{\rm vir}$ in the Hydro Rot case. Gas within the MHD Rot's halo maintains these high temperatures over a longer duration, approximately a Gyr longer than in the case of no MHD. %In the MHD Rot run, this slight additional thermal pressure from a hotter CGM, as well as the additional magnetic pressure, helps the ambient mass to buoyantly rise beyond thescp edsc virial radius. Mass moving from inside 12 kpc to outside seems to be more a consequence of gas that has too high angular momentum, thus it is kicked out of the center, extending the disk at late times, not necessarily heating from turbulent stiring. Especially because the turbulent stirring is constant in time. 

Both runs appear to come out of their cooling minima after nearly 4~Gyrs of evolution. Soon after, we also see the mass between 12~kpc and $R_{\rm vir}$ rise in the Hydro Rot halo. This is largely due to the ejection of material from the central region and somewhat from the re-accretion of material outside $R_{\rm vir}$. The ejection of material from the center that causes the rise in mass seen in this region is observed 2~Gyrs later in the MHD Rot halo as the magnetic field helps to keep cold gas contained to the center for longer periods of time. Furthermore, these runs maintain their increased masses in this interim region for the remainder of their evolution. The mass-weighted average temperature trends approach a minimum, with log~$T$ being $\approx$~5.6 in the MHD Rot halo and 5.5 in the Hydro Rot halo prior to rising for the remainder of the simulation.

\subsection{CGM Structure} 
\begin{figure*}
    \includegraphics[width=1.0\linewidth]{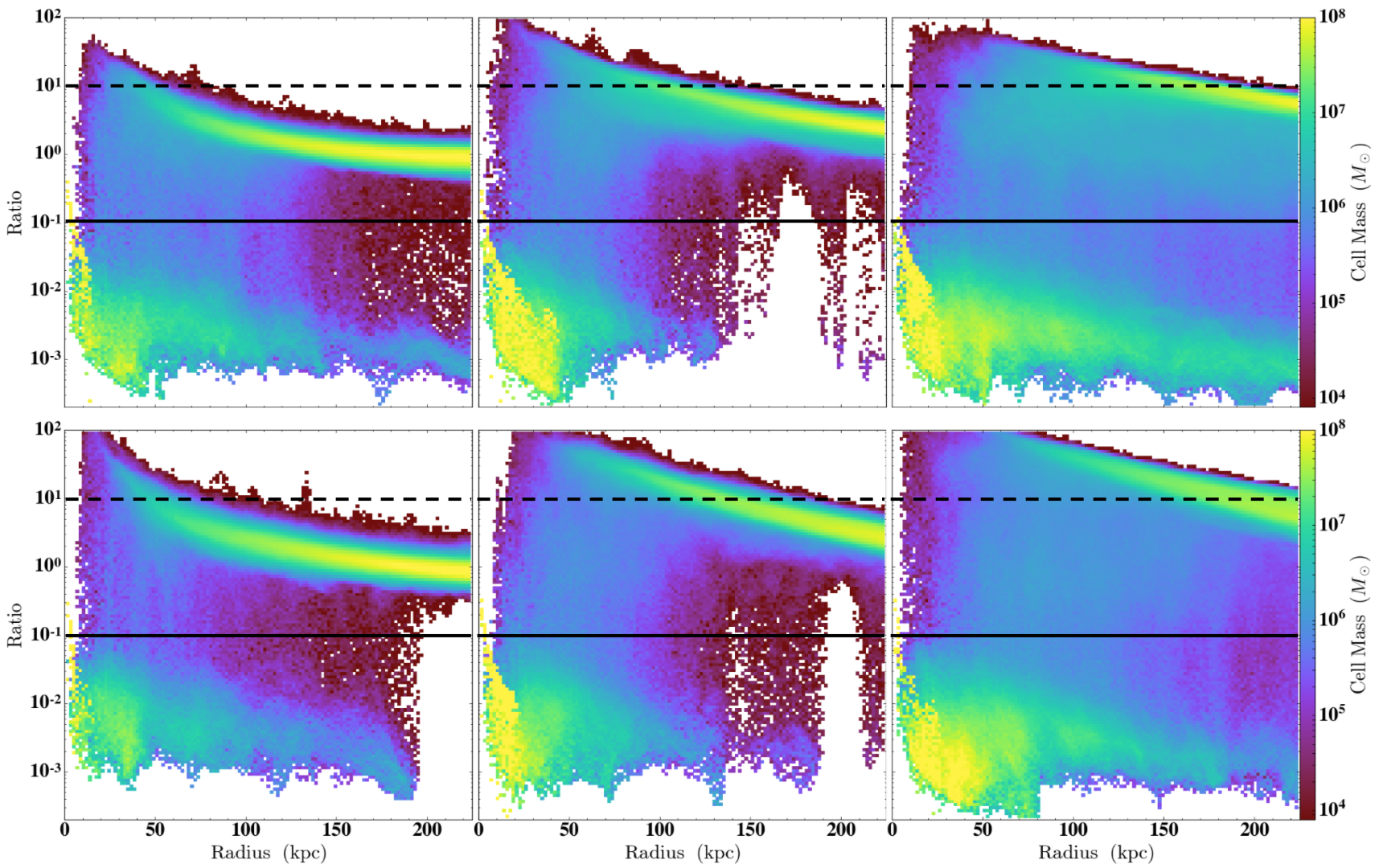}
    \caption{$t_{\rm cool}$/$t_{\rm ff}$ ratios vs. radius showing the total mass in a bin for the Hydro Rot (top) and MHD Rot (bottom) runs. We specifically show these at 3 (left), 6 (middle), and 9 (right) Gyrs. Plots are made using 128 bins for the x and y quantities. For reference, we include a black dashed line to show this ratio at 10 and a solid line to show this at 10$^{-1}$.}
    \label{fig:ratio_mhd_rot}
\end{figure*}

In Figure~\ref{fig:dens_mhd_rot}, we show the radial profile of the number density, $n$, at 3, 6 and 9~Gyrs. Gas is found at a variety of densities during these times, with most of the mass tracing the diffuse ambient medium and dense structures, such as clouds and filaments. The Hydro Rot run tends to have a wider range of densities, and at 6~Gyrs, it shows a large amount of mass with $n >$~10$^{-2}$~cm$^{-3}$ spread across the inner 50~kpc of the halo. In the MHD case, however, mass consolidates to smaller radii, reaching densities above 10$^{-1}$~cm$^{-3}$ along with possessing mass beyond 125~kpc with $n \approx$~10$^{-3}$~cm$^{-3}$. 

After 6~Gyrs, the halos become very dynamic with angular momentum transfer driving dense, cold outflowing gas throughout, some of which travels beyond $R_{\rm vir}$ by 9~Gyrs, while also having infalling, cooling gas. Both raise the density in the ambient medium by 9~Gyrs. Also at this time, we can see that both runs have moved their mass outward, with less of it having $n \gtrsim$~10$^{-1}$~cm$^{-3}$ in the inner 50~kpc and more with 10$^{-2} < n <$~10$^{-4}$~cm$^{-3}$ spread throughout the halos. 

%We also find that gas in the MHD Rot run above and below the disk quickly becomes ambient halo densities much faster than it does in the Hydro Rot halo.  with gas being near $T_{\rm vir}$ beyond $\approx$~100~kpc. itself near hen looking at these densities as a function of scale height, we find that the MHD Rot halo shows a steep decline in densities beyond $\approx$ 100~kpc at each of these times, as gas above and below the disk is near $T_{\rm vir}$ with densities below $\approx$~10$^{-28}$~g~cm$^{-3}$. This is in comparison to the Hydro Rot halo which instead maintains cool, $T \lesssim$~10$^{4}$~K, gas with densities $\approx$~10$^{-26}$~g~cm$^{-3}$ out to $r \approx$~175~kpc. In the plane of the disk, however, gas shows a gradual decrease in density and temperature with average trends that minimally vary between the runs.

%These trends have slightly changed by 6~Gyrs, as illustrated in the central column of Figure~\ref{fig:dens_cell_mass_compare_all}. Here we see that gas at $r \gtrsim 60$~kpc in the MHD Rot run has slightly higher densities at this time as compared to the Hydro Rot run, and this additional dense gas found is to be orbiting within the plane of rotation. Both runs show a more gradual decrease in the mass-weighted density when plotted as a function of planar distance, with the MHD Rot halo having slightly denser gas at more distant radii, along with a sharp decline in these densities when plotted as a function of vertical distance from the central galaxy. 

In Figure~\ref{fig:ratio_mhd_rot}, we examine the radial profile of the ratio of cooling times to free fall times, $t_{\rm cool}$/$t_{\rm ff}$, for these runs at 3, 6, and 9~Gyrs where 
\begin{equation}
t_{\rm cool} \equiv \frac{E_{\rm in}}{\sum_i \Lambda_i (T,Z,n_i)}, 
\end{equation}
\begin{equation}
t_{\rm ff} \equiv (2r/g)^{1/2},
\end{equation}
$E_{\rm in}$ is the internal energy, and $i$ denotes the $i-th$ ion. 

This ratio gives insight into the self-regulated balance maintained between gas condensing out of the ambient medium and feedback from the host galaxy in the form of an active galactic nuclei (AGN), star formation, and/or supernovae. Once triggered, such feedback may prevent further condensation from the ambient halo \citep[e.g.][]{1977ApJ...211..638S,gaspari2012mechanical, mccourt2012thermal, 2015voit, voit2017global}. 

The origin of this ratio arises from early theoretical work which attemped to understand thermal instabilities in the galactic context \citep[e.g.][]{1953ApJ...118..513H,1965ApJ...142..531F,1977MNRAS.179..541R}. When this ratio becomes too low (i.e. $t_{\rm cool}$/$t_{\rm ff} \lesssim 1$), gas may condense out of the halo and rain onto the central galaxy. As gas precipitates, any one of the aformentioned energetic feedback processes may come into play which raises this ratio in the ambient halo to $\approx 10$. This cycle facilitates the formation of a multiphase medium around galaxies, especially near the central regions of the CGM.

\begin{figure*}
    \includegraphics[width=1.0\linewidth]{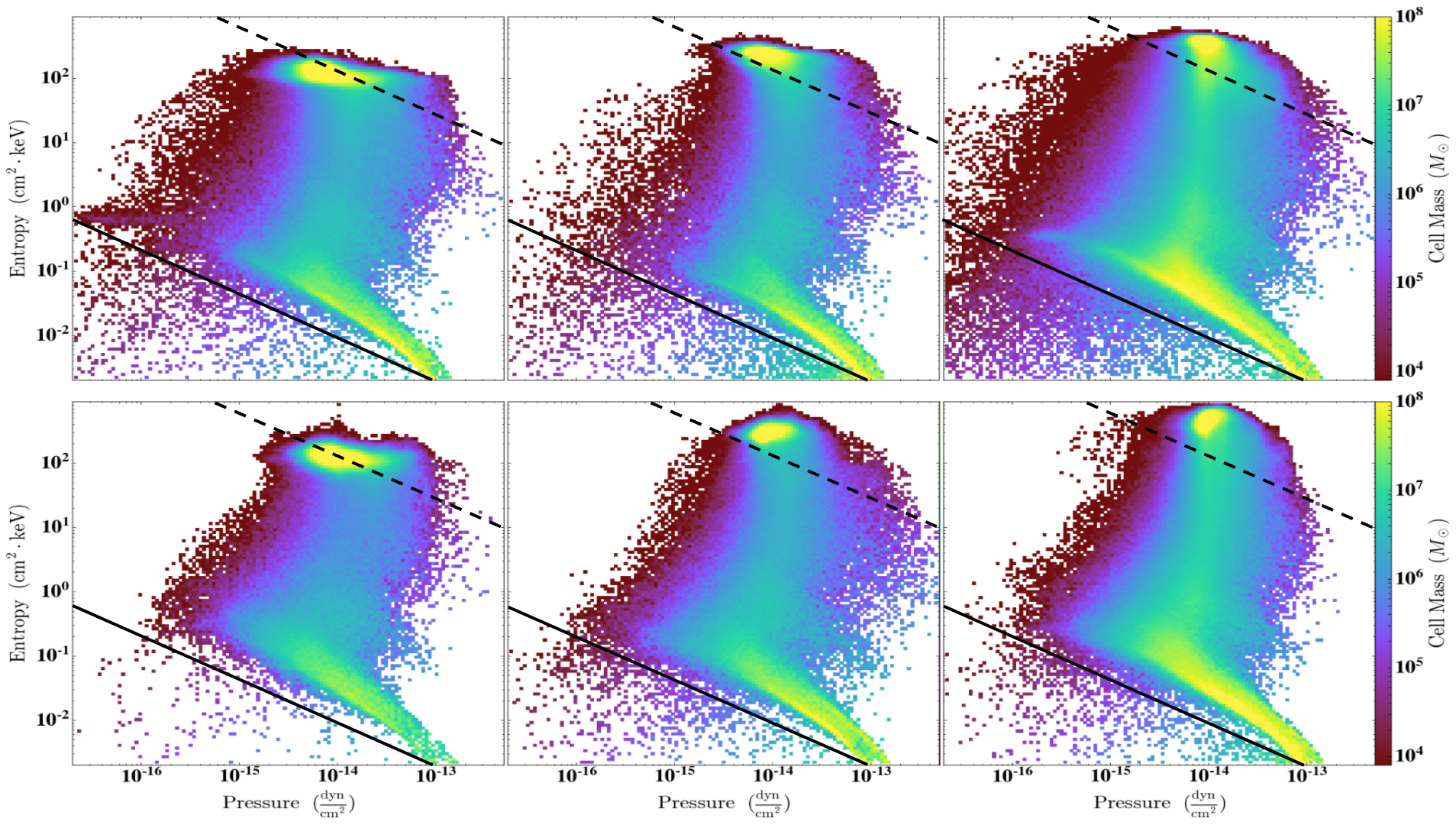}
    \caption{Entropy vs. thermal pressure showing the total mass in a bin for the Hydro Rot (top) and MHD Rot (bottom) runs. We specifically show these at 3 (left), 6 (middle), and 9 (right) Gyrs. Plots are made using 128 bins for the x and y quantities. For reference, we include a black dashed line to show the entropy-pressure profile at $T_{\rm vir}$ and a solid line to show this at 5000~K.}
    \label{fig:entropy_mhd_rot}
\end{figure*}

Others have conducted simulations of the CGM, testing different values of $t_{\rm cool}$/$t_{\rm ff}$ $\lesssim$~10 and their impacts on the development of a multiphase medium \citep[e.g.][]{mccourt2012thermal,2012MNRAS.420.3174S,2017MNRAS.466..677G,2019MNRAS.488.3195C,stern2019cooling}, however \citet{esmerian2020thermal} produced a bimodal distribution for these ratios in their Feedback in Realistic Environments (FIRE) simulations of Milky Way-like galaxies. Looking at our halos, we find a similar bimodal distribution for $t_{\rm cool}$/$t_{\rm ff}$, resembling what was found in the \citet{esmerian2020thermal} simulations. The hot and diffuse ambient medium lies between 1 - 100, while the other traces $T \lesssim 10^{5}$ gas and lies between 10$^{-4}$ - 10$^{-1}$. We also find a slightly wider distribution of $t_{\rm cool}$/$t_{\rm ff}$ ratios in the MHD Rot case at 9~Gyrs. This run possesses gas with ratios beyond 100 that traces low densities around 10$^{-5}$~cm$^{-3}$, lying above and below the central disk-like structure.

We find that earlier in the simulations, this extraplanar gas (not shown), is hot with temperatures $\gtrsim T_{\rm vir}$ and $t_{\rm cool}$/$t_{\rm ff}$ ratios $\gtrsim$~10. This is likely a consequence of the turbulent stirring, that we set to be strongest near the center, along with this region not having the additional cooling that results from the further mixing brought on by the co-rotation of gas. After $\approx$ 6~Gyrs, densities in these regions increase in the Hydro Rot run. During this stage, matter falls towards the center from above and below the central disk-like structure. This differs from the MHD case in which infalling matter has more angular momentum and thus funnels through the $y=0$ plane, within the extended disk. At 9~Gyrs, these regions of cooler gas above and below the disk in the Hydro Rot run extends much farther as compared to the MHD case. 

Lastly, we examine the entropy and thermal gas pressure profiles, again, at 3, 6, and 9~Gyrs in Figure~\ref{fig:entropy_mhd_rot}. At 3~Gyrs, the ambient medium possesses a maximum $K \approx$~100~cm$^{2}$~keV in both runs which grows to about 350 and 500~cm$^{2}$~keV by 9~Gyrs in the Hydro and MHD Rot halos respectively, meaning that unless it is mixed into colder material, its cooling time is much longer than a Hubble time \citep{2002ApJ...576..601V,2003MNRAS.342..664O}. We also find the additional heating that results from the continuous stirring and initialized co-rotation previously mentioned in $\S3.1$ causes the ambient medium in both runs to be more constrained to $\approx$~10$^{-14}$~dyn~cm$^{-2}$ by 9~Gyrs. We further find the MHD Rot run possesses a pressure profile that spans a smaller range as compared to the non-MHD case at all times. This may also be seen in gas transitioning between the cold and hot phases with entropy values of 0.1 - 10~cm$^{2}$~keV.

\begin{figure}
    \includegraphics[width=1.0\linewidth]{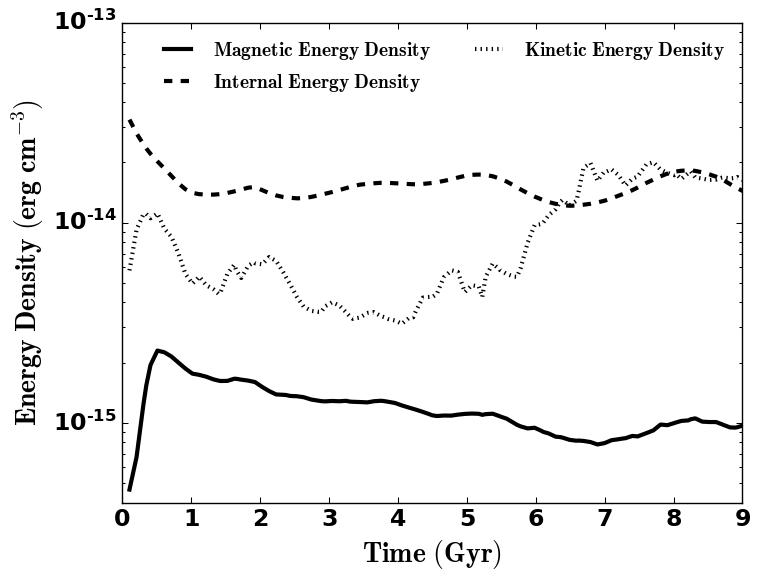}
    \caption{Volume-weighted average magnetic energy density, $u_{\rm B}$ (solid), internal energy density (dashed), and kinetic energy density (dotted) in the CGM gas between $12 < r < R_{\rm vir}$ vs. time for the MHD Rot run.}
    \label{fig:mag_mass_evo}
\end{figure}

\begin{figure*}
    \includegraphics[width=1.0\linewidth]{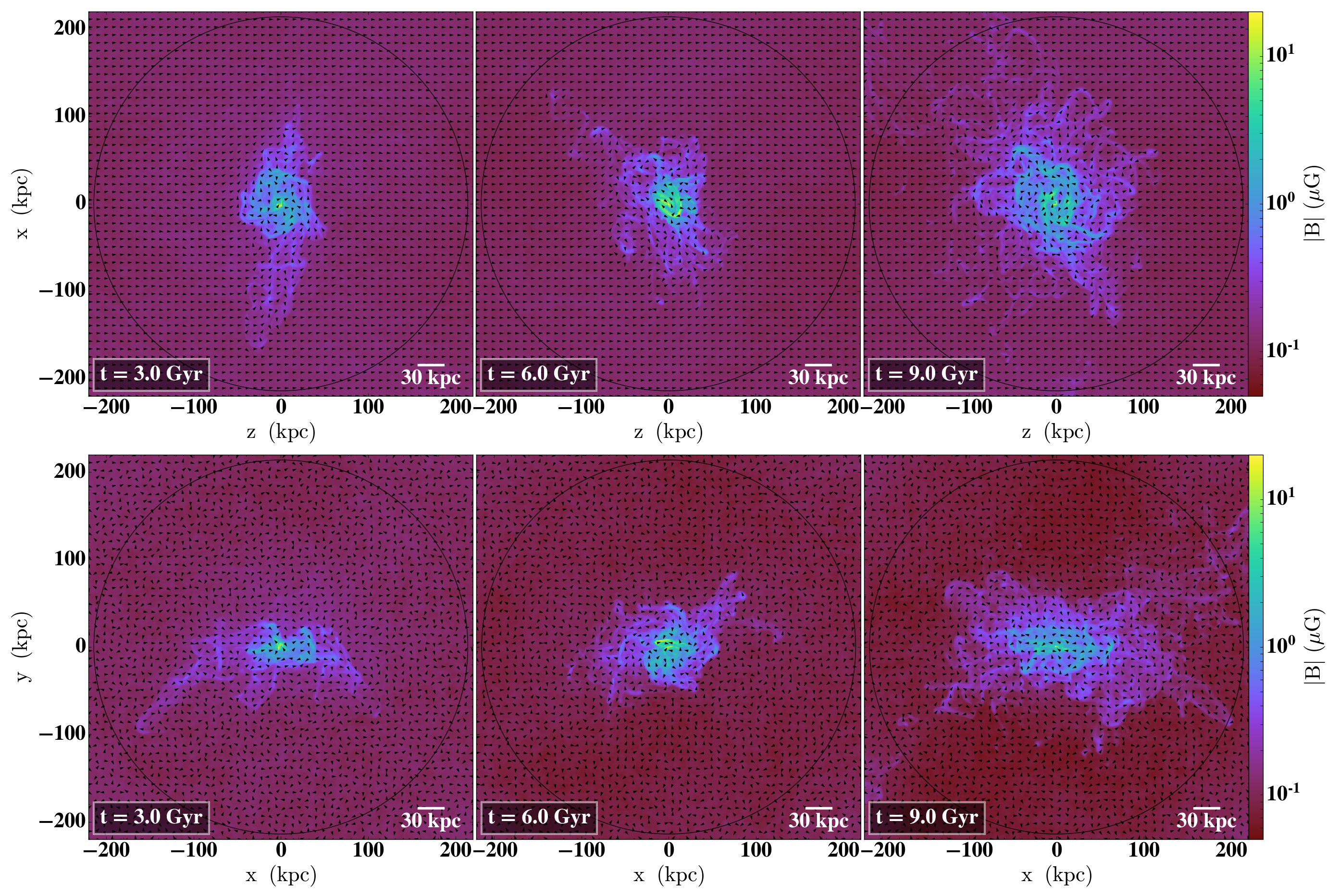}
    \caption{Density-weighted projections of $|B|$ at 3 (left), 6 (middle), and 9 (right)~Gyrs for the MHD Rot run. Projections through the halo from face-on and edge-on perspectives are shown in the top and bottom rows respectively. A black circle shows the virial radius at $R_{\rm vir} \approx 220$ kpc along with black quivers showing the direction of the normalized magnetic field vectors.}
    \label{fig:mag_structure}
\end{figure*}

Cooling and heating gas also lies across a nearly constant pressure range in both runs. We find that gas with $T \lesssim$~10$^{5}$~K possesses entropies < 1~cm$^{2}$~keV, and has a wider range of pressures between 10$^{-16}$ - 10$^{-13}$~dyn~cm$^{-2}$. In both runs, some of the mass is able to cool beneath our temperature floor at 5000~K, more of which is seen in our Hydro Rot run. This is where molecular cooling may become dynamically relevant, however this is not included in our simulations. By 9~Gyrs, both runs show lower entropy gas with $K \lesssim$~0.1~cm$^{2}$~keV that may be found at lower pressures below 10$^{-15}$~dyn~cm$^{-2}$. Further analysis shows us that this low thermal pressure gas has temperatures ranging from $\approx$~10$^{4-5}$~K as well as moderate densities around 10$^{-3}$~cm$^{-3}$. Lastly, we see the ambient medium possesses a nearly isentropic entropy profile at all times in both of our runs, shown in Figure~\ref{fig:entropy_radial} in the Appendix.

Finally, the continuous turbulent stirring along with the initialized co-rotation also raises the pressure in the medium beyond $R_{\rm vir}$ (not shown). We find that pressures in the medium beyond $R_{\rm vir}$ fluctuate between 6 - 9 $\times$~10$^{-15}$~dyn~cm$^{-2}$, slightly lower than those found in the hot component within the halo. As gas re-enters the box, it interacts with buoyantly rising gas between 300 and 400~kpc. The periodic fluctuation in pressure beyond $R_{\rm vir}$ is not seen in the gas inside $R_{\rm vir}$. It is unclear if these pressures outside of the virial radius have any influence the observable features within it, however, we do observe a slight increase of the temperature in this distant gas. 

\subsection{Magnetic Fields}

Next, we examine the magnetic field in the MHD Rot run and show the magnetic, internal, and kinetic energy densities as a function of time in Figure \ref{fig:mag_mass_evo}. We remind the reader that this run began with a seed $B_{\rm z} =$ 0.1$~\mu$G which corresponds to $u_{\rm B}=B^2/2\mu_{\rm 0}=4 \times 10^{-16}$ erg cm$^{-3}$ and a plasma $\beta$ of $\approx$ 170.

\begin{figure*}
    \includegraphics[width=1.0\linewidth]{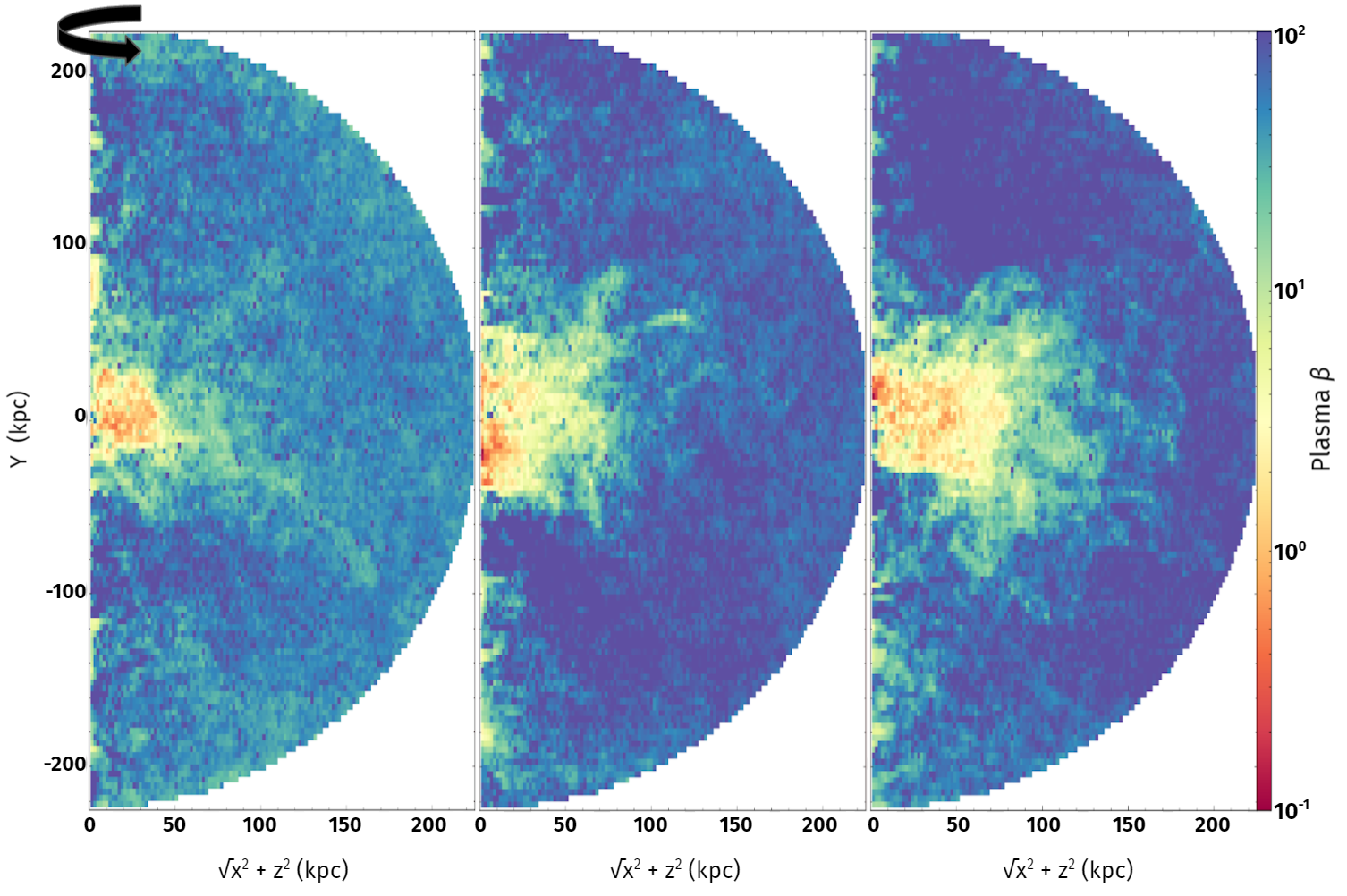}
    \caption{Height vs. $\sqrt{X^2 + Z^2}$ maps showing the mass-weighted average plasma $\beta$ in a bin at 3 (left), 6 (middle), and 9 (right)~Gyrs for the MHD Rot run. Plots are made using 128 bins for the x and y quantities.}
    \label{fig:plasma_beta}
\end{figure*}

The magnetic energy density grows very rapidly as the turbulent gas cools towards the center, amplifying $u_{\rm B}$ to nearly 5 times its initial value within 500~Myr. For comparison, $u_{\rm B}$ is able to reach $\approx$~12~\% of the internal energy density and $\approx$~35~\% of the kinetic energy density in the CGM at this time. Following this, $u_{\rm B}$ saturates until $\approx$~7~Gyr where it settles to about 5\% of the kinetic energy density for the remainder of its evolution. $|B|$ also fluctuates about 0.12~$\mu G$ at this time. 

As before, we take a closer look at the simulation at 3, 6, and 9~Gyrs, opting to show the density-weighted projections of $|B|$ from edge-on and face-on perspectives at these times in Figure~\ref{fig:mag_structure}, overlayed with black quivers that indicate the normalized magnetic field vectors.

Higher magnetic fields strengths between 1-10~$\mu$G are found in denser structures such as the cool, extended disk-like gas structure centered at $y=0$ and cooling filaments found throughout the halo. We generally find $|B|$ decreases more steeply above and below the central disk-like gas structure rather than along the plane of it. $|B|$ strengths nearing a $\mu$G may be found out to $r \approx 100$~kpc at 3~Gyr, with these strengths extending to 150~kpc by 9~Gyrs as increased amounts of dense filaments and clouds are dispersed throughout the halo during this time. A time of 9~Gyrs corresponds to when the halo shows outflowing material along with inflowing cooling gas from beyond $R_{\rm vir}$. The outflowing gas is able to carry magnetic energy from the central region out to distant radii, further magnetizing the halo. Although higher $B$-field strengths may be found at more distant radii during this later time in the MHD Rot run's evolution, we observe minimal difference in the $|B|$ radial profile throughout its evolution.

\begin{figure*}
    \centering
    \includegraphics[width=1.0\linewidth]{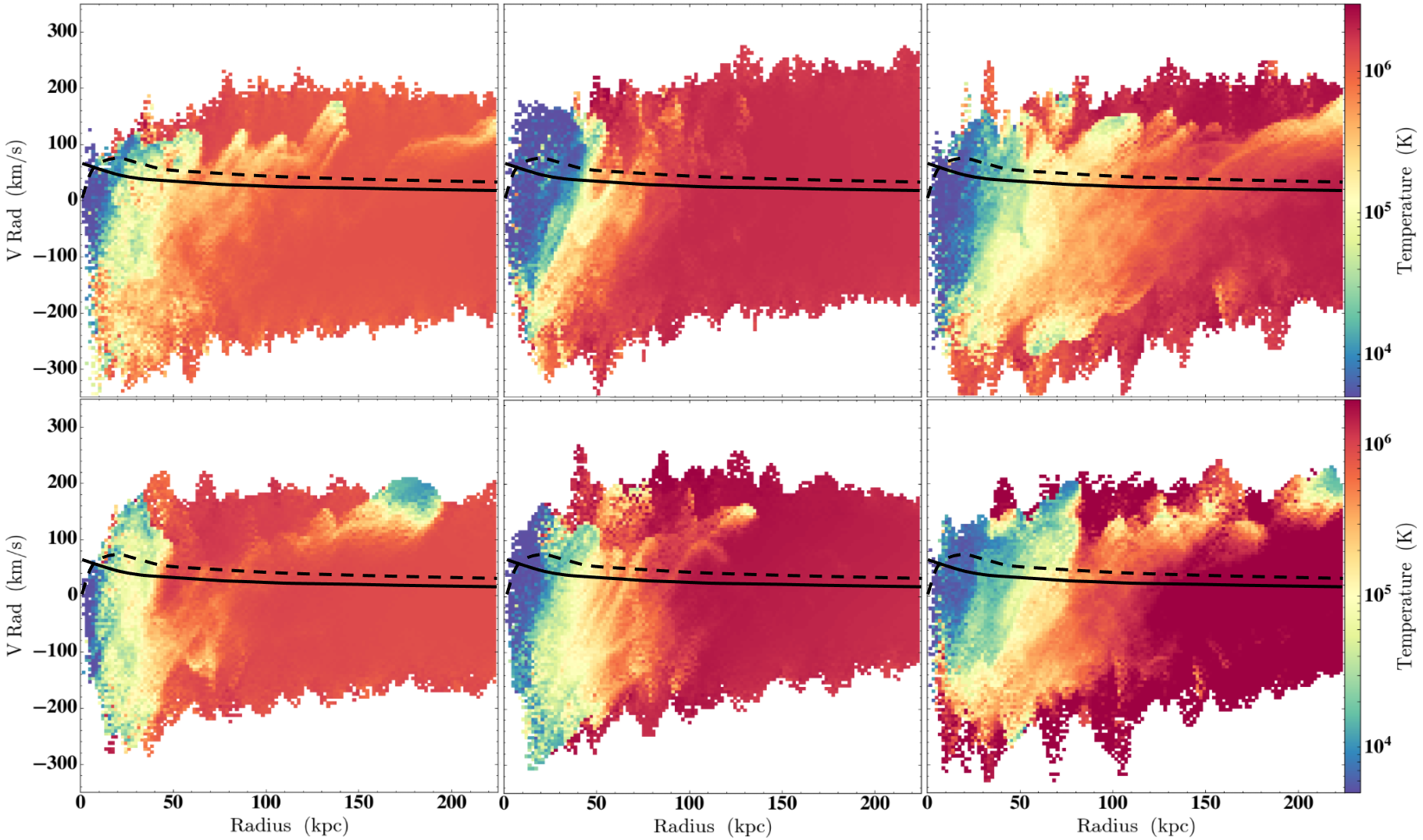}
    \caption{The mass-weighted average temperature as a function of radial velocity and radius for the Hydro Rot (top) and MHD Rot (bottom) runs at 3 (left), 6 (middle), and 9 (right) Gyrs. Plots are made using 128 bins for the x and y quantities. For reference, we include a solid black line to show the amplitude of the turbulence and a black dashed line to show the amplitude of the initial v$_{\rm rot}$ at y = 0.}
    \label{fig:radial_temp_compare}
\end{figure*}

In Figure~\ref{fig:plasma_beta} we show the distribution of average $\beta$ as a function of height and $\sqrt{x^2 + z^2}$ distance. As the halo evolves, the thermal pressure associated with the ambient medium increases as it heats and becomes more diffuse. This is shown by the increased volume with $\beta \approx$~100 while the gas with $\beta$ values $\lesssim$~1 is largely found within the extended disk-like structure at all times. Interestingly, gas above and below this structure and beyond 40~kpc, is mostly hot and diffuse, however, also possesses pockets where it is equally supported by thermal and magnetic pressures. These may be colder voids as higher temperatures would lead to higher $\beta$'s.

The gas with $\beta$ values $\lesssim$~1 may be susceptible to magnetic draping, an effect in which a gas cloud moving through a magnetized plasma is able to rapidly build up a magnetic layer which may shield it from developing instabilities \citep[e.g.][]{semenov1980structure,2008ApJ...677..993D,2020ApJ...892...59C}. When looking at the quivers in Figure~\ref{fig:mag_structure}, we indeed see magnetic fields that have oriented themselves to be predominately parallel to the surface of the cool structures such as the extended central disk-like structure. We speculate that this allows for the persistence of the extended disk-like structure at later times as mixing directly above and below it is inhibited.

Furthermore, the gas that we observed to have low thermal pressures in Figure~\ref{fig:entropy_mhd_rot} tends to be supported by a comparable amount of magnetic pressure as compared to thermal pressure. In fact, we find the gas with $K \lesssim$~0.1~cm$^{-2}$~keV typically corresponds to gas with $\beta$ values $\lesssim$~1.

\subsection{Kinematics \& Ions}

Finally, we compare the kinematics and distributions of ions in the two runs, the properties that are most often constrained by observations. In Figure~\ref{fig:radial_temp_compare}, we show the mass-weighted average temperature as a function of radial velocity, $V_{\rm radial}$ and radius at 3, 6, and 9~Gyrs, overlaid with the amplitude of turbulence and initial v$_{\rm rot}$ at $y=0$ for reference.

In general, we find a smaller spread in radial velocities for the MHD case as the magnetic field seems to restrict the most extreme motions of the gas. We see that velocities are, on average, $\approx$ 10~km~s$^{-1}$ lower than those found in the Hydro Rot run. We can also see how the MHD Rot halo progresses towards hotter temperatures, leading to a difference in the cooling that develops at late times. Both runs develop cool, typically dense outflowing gas structures that originate from the central region at 9~Gyrs, albeit more pronounced in the MHD Rot case. This is due to the combination of turbulent driving, which is the only source of feedback in these simulations, and angular momentum transfer which helps to expel cool gas from the central region.

\begin{figure*}
    \includegraphics[width=1.0\linewidth]{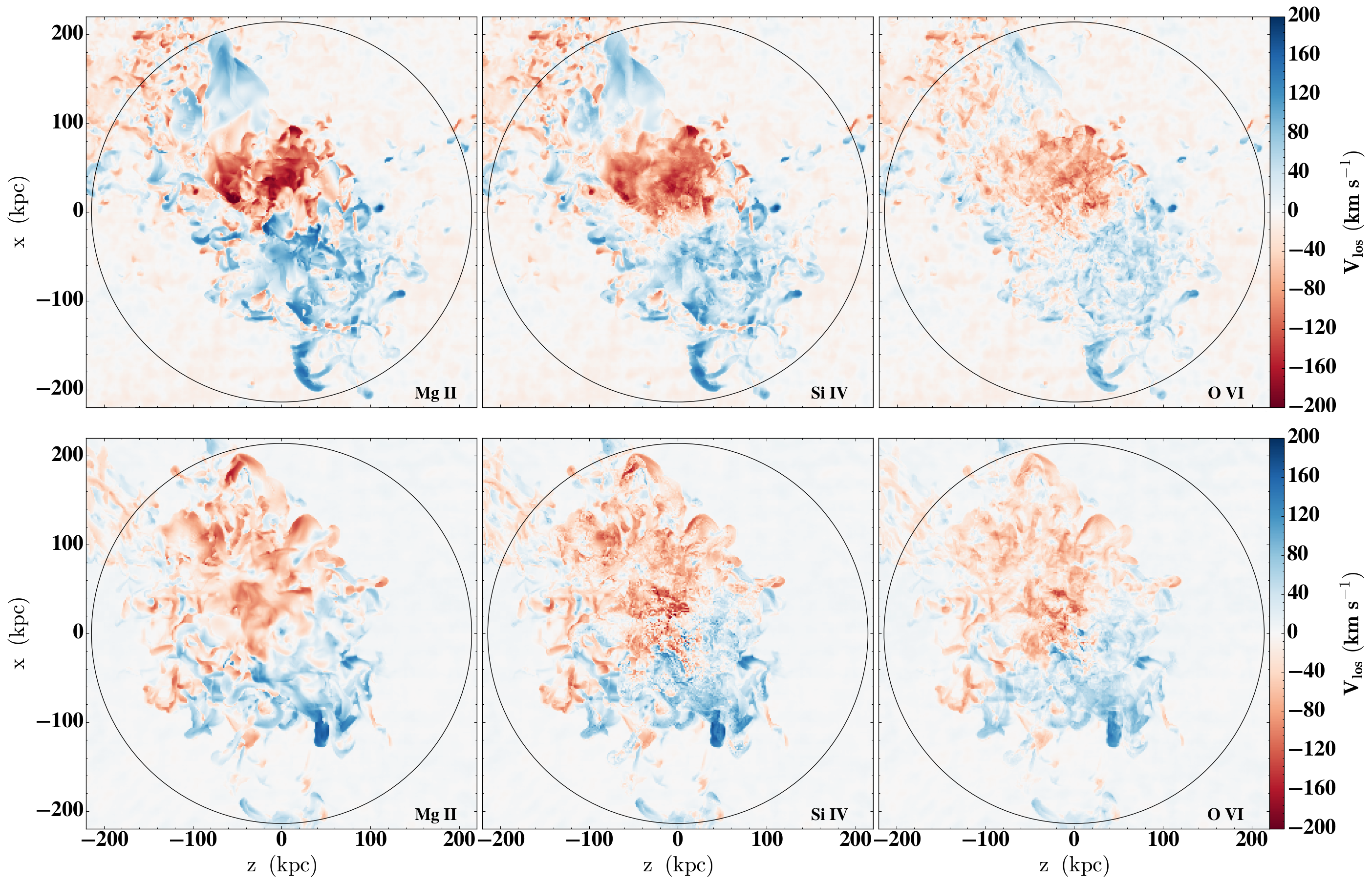}
    \caption{\ion{Mg}{2} (left), \ion{Si}{4} (middle), and \ion{O}{6} (right) weighted line-of-sight velocity projections for the Hydro Rot (top) and MHD Rot (bottom) runs. Projections are from a face-on perspective at 9~Gyrs. A black circle shows the virial radius at $R_{\rm vir} \approx 220$ kpc.}
    \label{fig:v_los_face}
\end{figure*}

Figure~\ref{fig:v_los_face} shows ion-weighted projections of the line-of-sight velocities for these runs, illustrating how the MHD influences the motions of commonly-observed ions. From the face-on perspective shown in this plot, we do not have to consider the overall co-rotation velocity component. Here we find the hot component tracing the ambient medium to be largely static, with low line-of-sight velocities near zero while the gas within cooling structures traces the fastest moving material. In particular, the low and intermediate ions, \ion{Mg}{2} and \ion{Si}{4}, show the largest line-of-sight velocities, approaching $\pm$200~km~s$^{-1}$ in the dense cores of cooling structures while the high ion phase, as traced by \ion{O}{6}, shows smaller velocities overall. The kinematics of the gas is a combination of turbulent motions from the consistent driving, thermal motions from its inherent temperature, and motions due to the gas residing in a gravitational potential.

In B20, inhomogeneous turbulent stirring was made to be strongest near the center to mimic the star-forming galaxy environment with the majority of its feedback resulting from the center, as well as the impact of gas accreted from the intergalactic medium, which is likely to deposit the most energy per unit volume near the center of the halo. This stirring profile promoted a convective flow as hot material was driven to distant radii and replaced by gas condensing out of the ambient halo. When lower degrees of stirring were tested, gas simply collapsed towards the center with very limited mixing between the cold and hot components.

In this study, however, co-rotation provides a natural mechanism for distributing gas and promoting interactions between cold and hot material. Therefore, a change in the amount of turbulent stirring, or in its radial profile by adjusting $\alpha$ in Equation~\ref{equ:trend}, may lead to different features arising. In the case of stronger stirring, one may observe more outflows and possibly more gas that has been collisionally ionized in the central region, while less stirring may result in less of these. In the case of inhomogeneous stirring that instead increased with radius, one might observe slightly hotter distant ambient gas, as the turbulence would eventually dissipate its energy as heat, along with an increased level of variance in densities found in the outer regions (which may produce changes in the resulting ion column densities in these regions). These considerations may be explored in a future study.

Although not shown here, we find gas moving at faster velocities in the MHD case as compared to the hydrodynamic case when the halos are projected through from an edge-on perspective. It may be the case that magnetic fields help to move gas through the central, $y=0,$ plane which results in faster motions of this gas.

Moreover, we find the fastest moving clouds also have $t_{\rm cool}/t_{\rm ff}$ ratios $\lesssim$~10$^{-1}$ meaning they are cooling at least 10x faster than they are falling, which could potentially lead to star formation in their cores \citep[e.g.][]{turner2015highly}. These simulations do not consider star formation (and have a maximum resolution of 1.6~kpc), however this may have implications higher resolution simulation that do consider the formation of star clusters.

\begin{figure*}
    \includegraphics[width=1.0\linewidth]{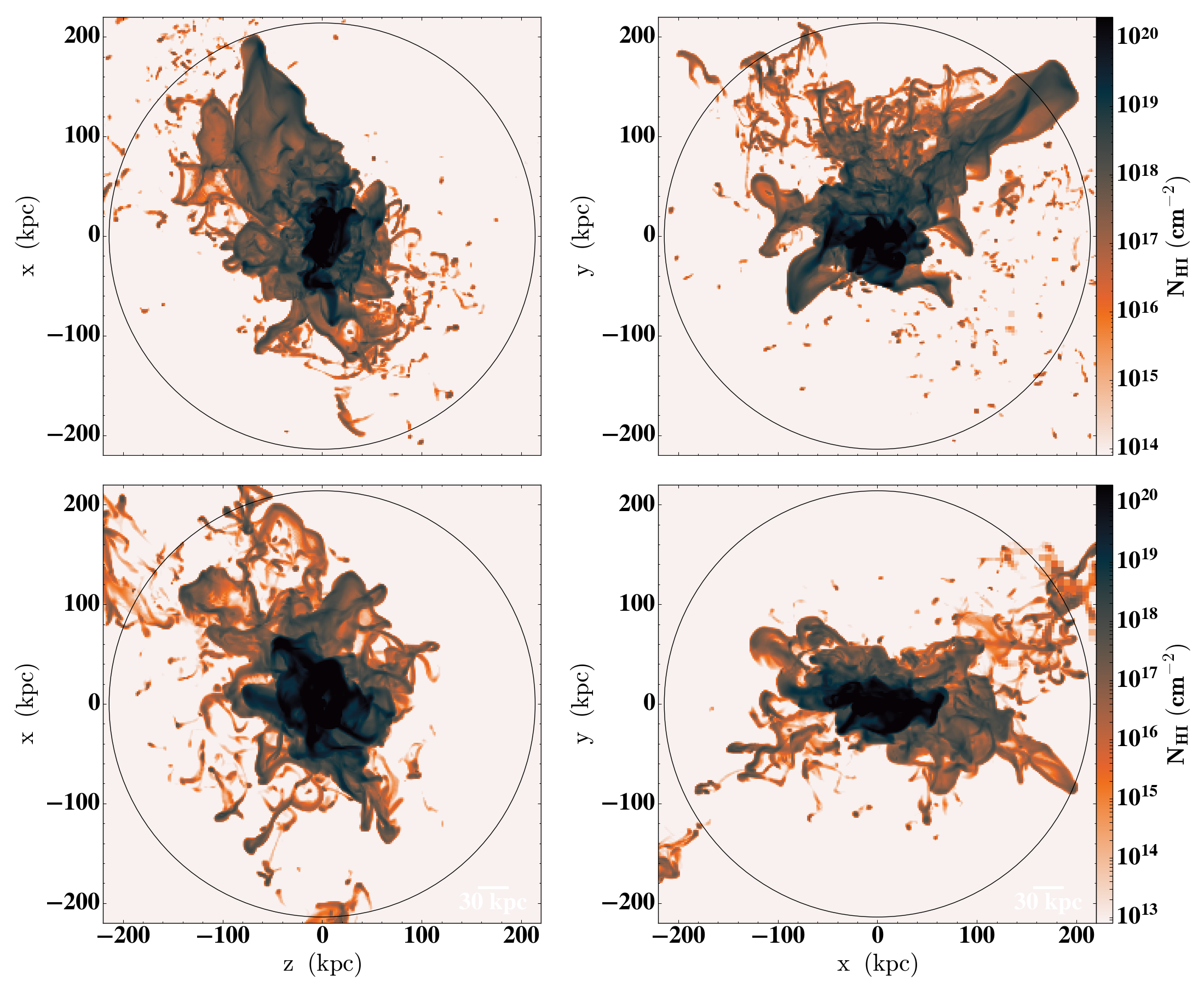}
    \caption{\ion{H}{1} column density maps for the Hydro Rot (top) and MHD Rot (bottom) runs at 9~Gyrs. Projections through the halo from face-on and edge-on perspectives are shown in the left and right columns respectively. A black circle shows the virial radius at $R_{\rm vir} \approx 220$ kpc.}
    \label{fig:hi_projs}
\end{figure*}

To further examine the similarities and differences between the cooling structures found in these two runs, we include the projected column density maps for \ion{H}{1} at 9~Gyrs from both the face-on and edge-on perspectives in Figure~\ref{fig:hi_projs}. Again, 9~Gyrs is a point in the evolution that is characterized by large-scale, fairly dense outflowing and inflowing gas clouds and filaments. We see that there are less cool clouds distributed throughout the MHD Rot run's CGM as compared to the hydrodynamic case. 

The smaller amount of cool clouds dispersed throughout the MHD halo may be a consequence of evaporation as the surrounding medium is hotter in this run as compared to the Hydro case. Furthermore, it has been shown that magnetic fields do not significantly increase the lifetime of cool clouds due to compression (in the case of tranverse fields) or the development of cloud tails (in the case of aligned fields), both of which promote mixing with the ambient medium \citep{2020ApJ...892...59C}. As both simulations evolve, we find an increasing lower limit of \ion{H}{1} columns as well as a covering fraction of 100\% for \ion{H}{1}, both seen within $\approx$~50~kpc. 

This high covering fraction is most likely a consequence of the idealized nature of our simulations as the amount of cool gas near the central region continues to build without additional ionizing sources of feedback, such as a star-forming UV background, AGN feedback, or cosmic rays. Also, our simulations do not consider radiative transfer, which would provide additional means for heating and possibly ionization. These relevant radiative backgrounds (i.e. star-forming UV background, AGN feedback, or cosmic rays) would provide additional heating to the central region of both halos, likely reducing some of the cooling that takes place in the central region, and even ionizing some of the dense \ion{H}{1} that accumulates there.

\begin{figure}
    \includegraphics[width=1.0\linewidth,height=0.8\textheight]{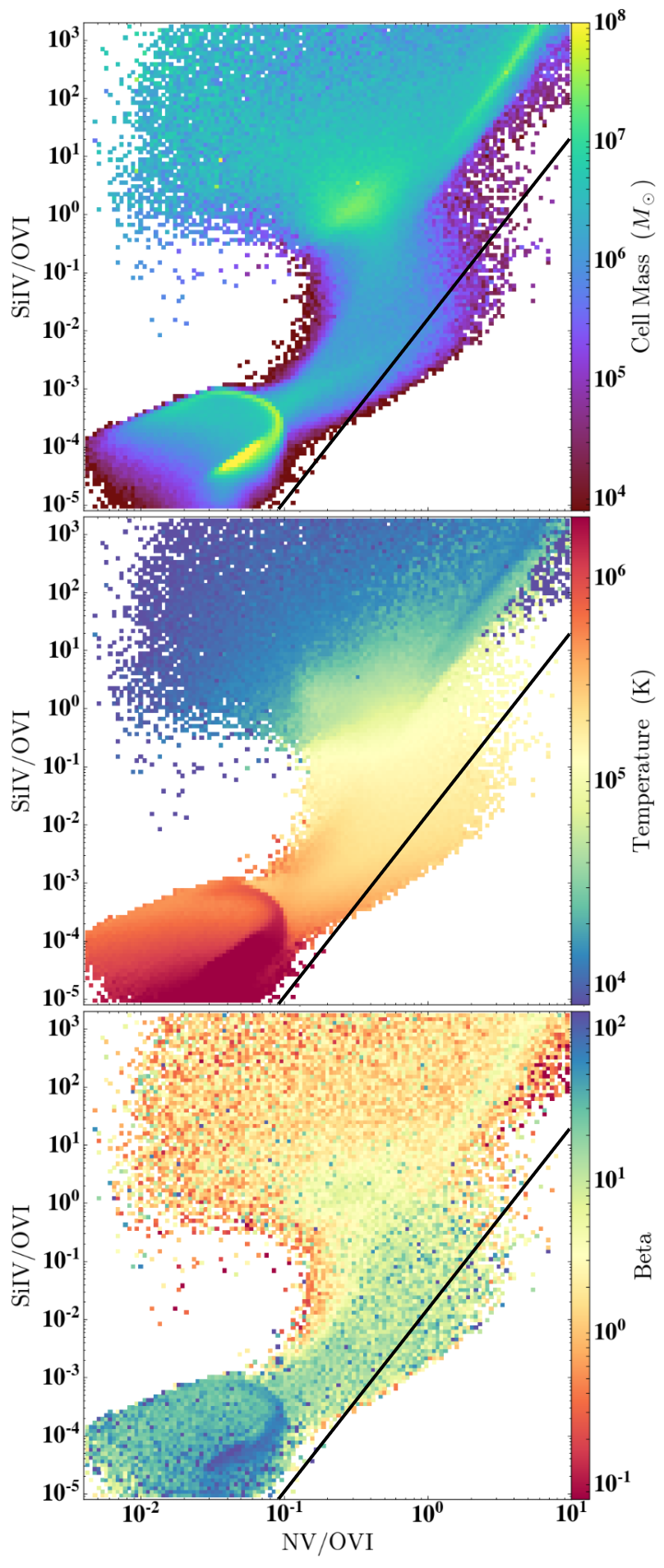}
    \caption{Panels showing the \ion{Si}{4}/\ion{O}{6} vs. \ion{N}{5}/\ion{O}{6} ratios with the total mass in a bin (top), mass-weighted average temperature (middle), and mass-weighted average $\beta$ from the MHD Rot run at 9~Gyrs. These are made using 128 bins for the $x$ and $y$ quantities. The black line shows a single-zone Cloudy model at 30\% solar metallicity photoionized by a redshift zero HM2012 EUVB with -4 < log~$U$ < -1 as a reference. This range in $U$ spans hydrogen densitites between 10$^{-1}$ and 10$^{-6}$ and yields temperatures between 10$^{4-5}$. COS-Halos detections and limits from \citet{werk2016ApJ...833...54W} would be located in the $T$ = 10$^{5-6}$~K region of these plots.}
    \label{fig:si4_o6_compare}
\end{figure}

In Figure~\ref{fig:si4_o6_compare}, we show the \ion{Si}{4}/\ion{O}{6} and \ion{N}{5}/\ion{O}{6} ratios for the MHD Rot run at 9~Gyrs and a single-zone CLOUDY model for reference. Although we do not show these ion ratios for the Hydro case, we find that both of the run's ion ratios look very similar to what is shown. Moreover, these ion ratios do not change shape throughout the 9~Gyr evolution, except for being slightly less dispersed at earlier times. In the MHD case, gas with \ion{Si}{4}/\ion{O}{6} ratios $\gtrsim$~1 has temperatures $\lesssim$~40,000~K and $\beta$ values $\lesssim 1$, whereas this gas has no magnetic pressure support in the hydrodynamic case.

We also performed a cut on the data to see whether the ratios change considerably if we consider ``realistic" outer CGM observations, i.e. those with \ion{O}{6} column densities above 10$^{13}$~cm$^{-2}$. We assumed a path length of 100~kpc to obtain a conservative lower limit on $n_{\rm OVI}$ of 3~$\times$~10$^{-11}$~cm$^{-3}$. This cut essentially removes the most diffuse ambient halo gas that would be unlikely to be observed in the spectra of background quasars. The removal of this gas, however, does not alter the overall shape of these distributions. 

The majority of COS-Halos detections and limits presented in Figure 12 of \citet{werk2016ApJ...833...54W} have \ion{Si}{4}/\ion{O}{6} $\lesssim 10^{-1}$ \ion{N}{5}/\ion{O}{6} $\lesssim 10^{-1}.$ While these are difficult to explain from equilibrium models, they naturally correspond to gas with $T$ between 10$^{5-6}$~K in our simulations. In the MHD Rot run, this region of parameter space also houses gas with $\beta$'s $\lesssim$~1, meaning some of these systems could trace gas in which the dominant pressure is magnetic.
% Many of these observations are upper limits due to non-detections of \ion{N}{5} which showed a preference in \ion{N}{5}/\ion{O}{6} leftward of 10$^{-1}$. 

\begin{figure*}
    \includegraphics[width=1.0\linewidth]{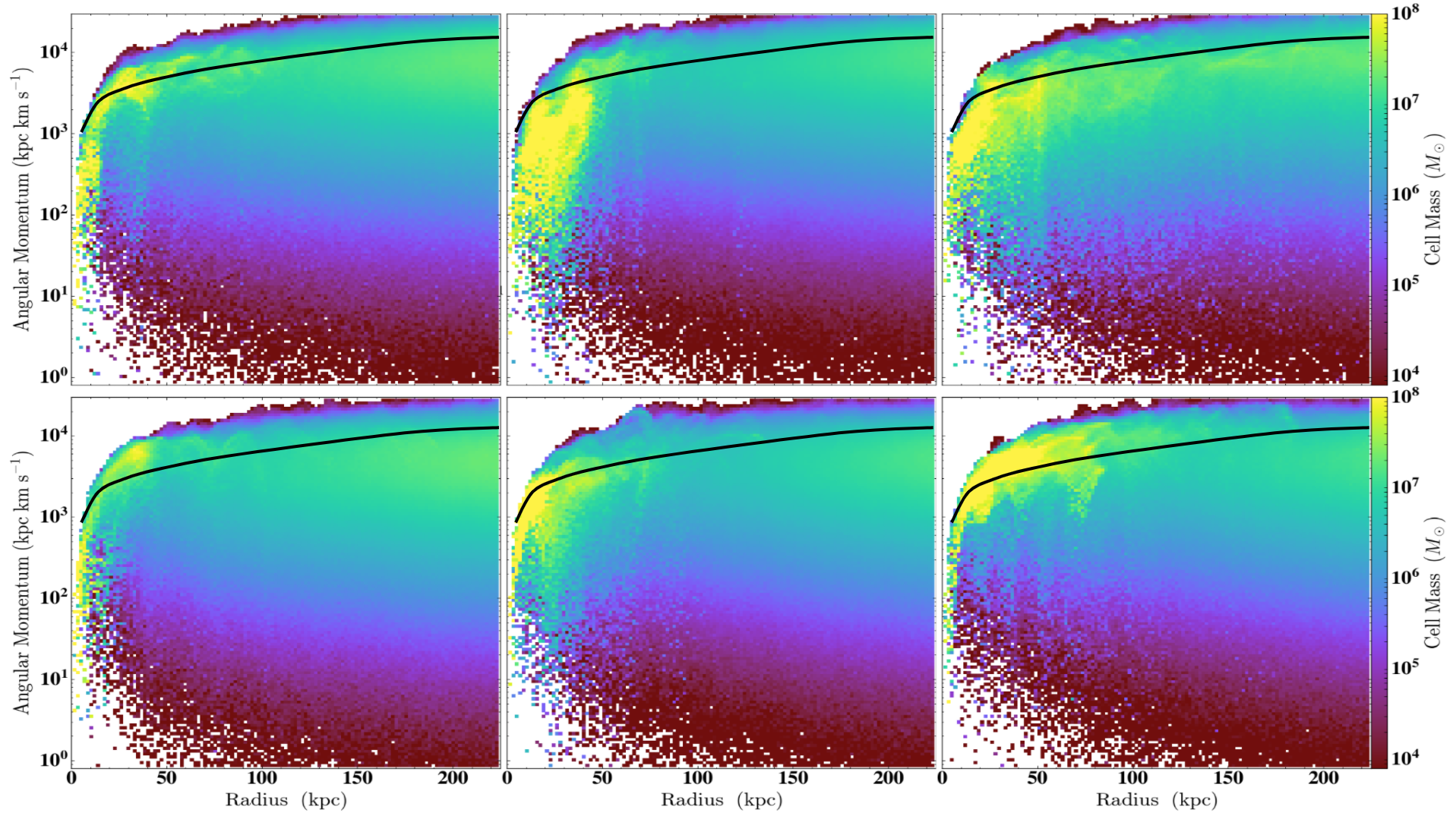}
    \caption{Distribution of CGM mass as a function of radius and angular momentum per unit mass, for the Hydro Rot (top) and MHD Rot (bottom) runs at 3 (left), 6 (middle), and 9 (right) Gyrs. These are made using 128 bins for the $x$ and $y$ quantities. For reference, we include a solid black line to show the specific angular momentum profile for Keplerian orbit.}
    \label{fig:angular_mhd_rot}
\end{figure*}

In Figure~\ref{fig:angular_mhd_rot}, we show the distribution of CGM mass as a function of radius and angular momentum per unit mass (otherwise known as the specific angular momentum). We further show a Keplerian specific angular momentum profile for comparison, which we will denote as $L_{\rm Kep}$. At 3~Gyrs, we find that the majority of the mass in the inner 50~kpc of the Hydro Rot halo lies along $L_{\rm Kep}$ and the majority of this mass in the MHD case is above the Keplerian profile. Such gas is under the influence of turbulent stirring along with co-rotation and this difference indicates larger velocities in the gas residing in the inner halo of the MHD run. This higher angular momentum gas is found near the plane of the central disk structure, centered at $y=0.$ At 6~Gyrs, we observe the majority of the mass in the inner 50~kpc has decreased in specific angular momentum in both runs. Being that the mass does not seem to have moved to larger distances, this may indicate a decrease in associated velocities. The mass in this inner region of the MHD halo lies along $L_{\rm Kep}$, while the mass in this same region of the Hydro Rot lies below it.

At 9~Gyrs, however, the velocities of the majority of the mass within 100~kpc increase again. At this time, the majority of mass in this inner region lies closer to $L_{\rm Kep}$ in the Hydro case and at or above the Keplerian profiles in the MHD case, indicating that it has been spun up at this later time in its evolution. Thus, it may be inferred that most of the mass within 100~kpc in the hydrodynamic case is traveling at smaller velocities as compared to the gas in the MHD central disk-like structure. Again, gas with the highest specific angular momentum is found along the plane of rotation at $y=0.$ We can also see this from the last two rows in Figure~\ref{fig:ndens_slices} which shows that in both of our runs the densest ($n \gtrsim 10^{-2}$~cm$^{-3}$) material containing most of the mass in the inner $\approx$ 100~kpc region lies within $\pm$~50~kpc of $y=0$.  %newfound cold material in the MHD Rot run finds itself with greater angular momentum than the cold gas in the Hydro Rot run. In the hydrodynamic case, cold gas is dispersed across a wide range of angular momenta, indicating that it may be found, not only in the extended disk, but also at a range of heights above and below it. This is in comparison to cold gas in the MHD Rot run which largely lies above 10$^{3}$~kpc~km~s$^{-1}$ in the inner 100~kpc of the halo, showing that it is more concentrated to within the extended disk structure. 

Being that this dense gas also houses the largest magnetic fields, it is likely that the magnetic fields help to facilitate this transfer of angular momentum throughout the cold gas. This transfer of angular momentum in the MHD case, occurs between 6 - 8~Gyrs, and is likely caused by the combination of large outflows of $v_{\rm r} \gtrsim$~100~km~$^{-1}$ material along with infalling condensate from the ambient medium that occurs predominantly along the central disk plane centered at $y=0$.

Finally, we note that \citet{2016ApJ...822...21H} found a spherically rotating model, rather than the thick disk model model adopted here, provided a good match to the observations of \ion{O}{7} absorption lines in the Milky Way halo. Adopting such a model would result in a slightly higher total angular momentum of the system, requiring a smaller $f_{\rm rot}$ in Equation~\ref{equ:rotation} to obtain a $\lambda$ near 0.03 for our halo. We speculate that spherical rotation may result in slight changes to the mixing and eventual cooling of hot gas, more so in the gas at the outer regions of the halo. However it is unlikely that these would lead to substantial changes in the structures, column densities, and other properties, especially since the scale heights of disk models that were compared to a spherical rotational model in \citet{2016ApJ...822...21H} were on the order of a few kpc, rather than the 50~kpc adopted here.

\section{Discussion \& Summary} \label{summary}

In this work, we describe the first simulation of the evolution of a magnetized Milky Way-like galactic halo that includes full non-equillibrium chemistry, rotation, and turbulence. Our idealized setup allows us to directly contrast this simulation with a hydrodynamic case in and draw conclusions on the impact of magnetic fields on the state multiphase CGM and its observational properties.

The addition of a co-rotational component enhances the mixing of hot and cold gas in both simulations, which leads to increased cooling. Halos in both simulations lose $\approx$ 14\% of their total baryonic mass due to the initial accretion shock and turbulent motions. At about 4 and 6 Gyrs for in the Hydro and MHD Rot runs respectively, mass begins to be ejected from the central 12~kpc region out into the CGM. The MHD halo also continues to lose mass from the ambient medium for the remainder of its evolution, leading to a somewhat less dense and hotter medium than in the hydrodynamic case. 

One consequence of the MHD Rot halo forming a hotter, more diffuse CGM, is that the gas above and below the extended disk attains longer cooling times with $t_{\rm cool}/t_{\rm ff}$ ratios > 100 by 9~Gyrs. Without MHD, gas is able to cool at more distant radii, which lowers the maximum $t_{\rm cool}/t_{\rm ff}$ ratios of extraplanar gas to < 100 as the densities in these regions increase. Both halos, however, develop bimodal $t_{\rm cool}/t_{\rm ff}$  distributions where one phase traces the hot component with 1 < $t_{\rm cool}/t_{\rm ff}$ < 100, and the other traces cold, dense gas with 10$^{-4}$ < $t_{\rm cool}/t_{\rm ff}$ < 10$^{-1}$. 

When examining the entropy-pressure phase space of our runs, we find that the MHD ambient medium rises to greater entropy throughout its evolution. Due to expansion, some of the lower entropy gas is able to cool below our temperature floor at 5000~K, the temperature below which radiative cooling is dominated by molecules, which are not included in our chemical network. We find that the MHD run keeps more gas above this temperature limit, and by 9~Gyrs, this case leads to a restriction in the pressure phase space of the intermediate, $T \approx$~10$^{5}$~K, gas to a narrower range as compared to the Hydro Rot run. Finally, gas in the MHD Rot run with $K \lesssim$~1~cm$^{2}$~keV also possess low $\beta$ values that indicate significant magnetic pressure support. 

We find ambient halo $B$-field strengths saturate to $\approx$~0.1~$\mu$G in the 9~Gyr evolution, back to the initial seed field strength. We also find higher strengths $\lesssim$~10~$\mu$G in denser structures that form in the inner halo. These denser structures typically show $B$-field lines that are oriented parallel to their surfaces, which may be an indication of magnetic draping that has become dynamically important. The presence of a magnetic field also helps enable the transfer of angular momentum throughout the cold gas. In the MHD case, we find a slightly more extended disk-like structure that forms in the central region and moves with faster velocities as compared to the hydrodynamic case in the later stages of evolution. This may also help to maintain the extended disk-like structure at late times as nearby gas outside of it interacts with it and gains angular momentum thereby keeping the nearby cooling and heating gas constrained to the disk plane.

Kinematically, we find that magnetic fields, overall, inhibits the radial motions of gas, with the MHD run showing a slightly smaller dispersion in radial velocities throughout its evolution. We find that dense structures showing outflowing and inflowing material, and the ions associated with them, are moving the fastest, at $v_{\rm r}$ upwards $\pm$~200~km~s$^{-1}$ in the Hydro Rot case. This material is observed to be moving somewhat slower in the MHD case when viewed from a face-on perspective. In both runs, it is the low, singly-ionized states, such as \ion{Mg}{2} found in the densest cores of the outflowing and inflowing gas, that typically possess the largest line of sight velocities, whereas the highest ionization states, such as \ion{O}{6}, move the slowest.

%As the MHD Rot halo continues to evolve, it is able to build regions of fairly dense \ion{H}{1} column densities that inhibits the formation of III and IV \textbf{ionization states of all elements} due to high recombination rates. These column densities grow to be, on average, 10 times larger than those found in the Hydro Rot halo, and this also yields larger column densities of doubly-ionized gas in the inner halo of the MHD Rot run. 
When looking at the \ion{Si}{4}/\ion{O}{6} and \ion{N}{5}/\ion{O}{6} ratios, we find very little difference between these runs, and furthermore, throughout their 9~Gyr evolution. One difference is that in the MHD Rot run, gas with \ion{Si}{4}/\ion{O}{6} $\gtrsim$~1 finds itself comparable or greater magnetic pressure support. %This parameter space appears to resemble the preferred parameter space of CGM observation \citet{werk2016ApJ...833...54W}, suggesting that much of these observations may be probing highly-magnetized, low $\beta$ material.

%Because the dense \ion{H}{1} is associated with the highest $B$-field strengths, this may have contributed to larger structures coming together in the MHD Rot case as cold gas continues to be produced through high recombination rates and protected from the influence of the hot ambient medium via magnetic draping. After 6~Gyrs, both halos enter periods of cooling wherein the Hydro Rot run cools down to a similar mass-weighted temperature as it began with, while the MHD case does so to a slightly higher temperature. Gas within the MHD Rot halo must find a way to cool from its higher temperature which leads to the low production of intermediate ions within about 100~kpc as these ions are short lived in the hot medium and either re-ionize or quickly cool to lower states. Outside 100~kpc, these column densities are comparable between the Hydro and MHD Rot runs. 

%These findings show the impact that magnetic fields may have on the evolution of the CGM, especially when it pertains to the inner halo gas as it allows for more dense \ion{H}{1} columns to develop wherein molecular cooling may become necessary to consider. 
Future work may include an expansion on our chemistry network to include dynamically relevant molecular cooling processes, as well as higher resolution to better resolve the densest structures that develop. This may also include varying the initial magnetic field strength to achieve higher $\beta$'s at initialization to observe how this may influence the saturation of the magnetic field. 

Moreover, we may investigate different turbulent feedback prescriptions (changing $\alpha$ in Equation~\ref{equ:trend}) to see if varying initial turbulent driving leads to very different results. Finally, including an additional star forming background in the inner CGM will allow us to gain a better understanding of how this radiative feedback would influence the large column densities that form around the central galaxy.

\section{Acknowledgments} \label{acknowlege}
We would like to thank Jessica Werk and Sarah Tuttle for their useful comments and feedback in the early stages of this work. We would also like to thank the referee for their detailed comments that greatly improved the manuscript. E.B.II was supported by the National Science Foundation Graduate Research Fellowship Program under grant No. 026257-001. The simulations presented in this work were carried out on the Stampede2 supercomputer at the Texas Advanced Computing Center (TACC) through Extreme Science and Engineering Discovery Environment (XSEDE) resources under grant TGPHY200071.\\

\textit{Software}: FLASH \citep[v4.5]{fryxell2000flash}, Cloudy \citep{ferland20132013}, yt \citep{2011turk}

\bibliographystyle{yahapj}
\bibliography{references}

\appendix

In this section we show the radial profile of entropy as a function of total mass for the Hydro and MHD Rot runs. These are shown so the reader may see where the majority of the mass lies and that the ambient medium is isentropic. Buoyancy therefore is unable to inhibit the progression of thermal instability into multiphase condensation. The centralized turbulent stirring creates a convective flow that progressively heats the entire atmosphere, facilitating the creation of the isentropic profile.

\begin{figure*}[h]
    \includegraphics[width=1.0\linewidth]{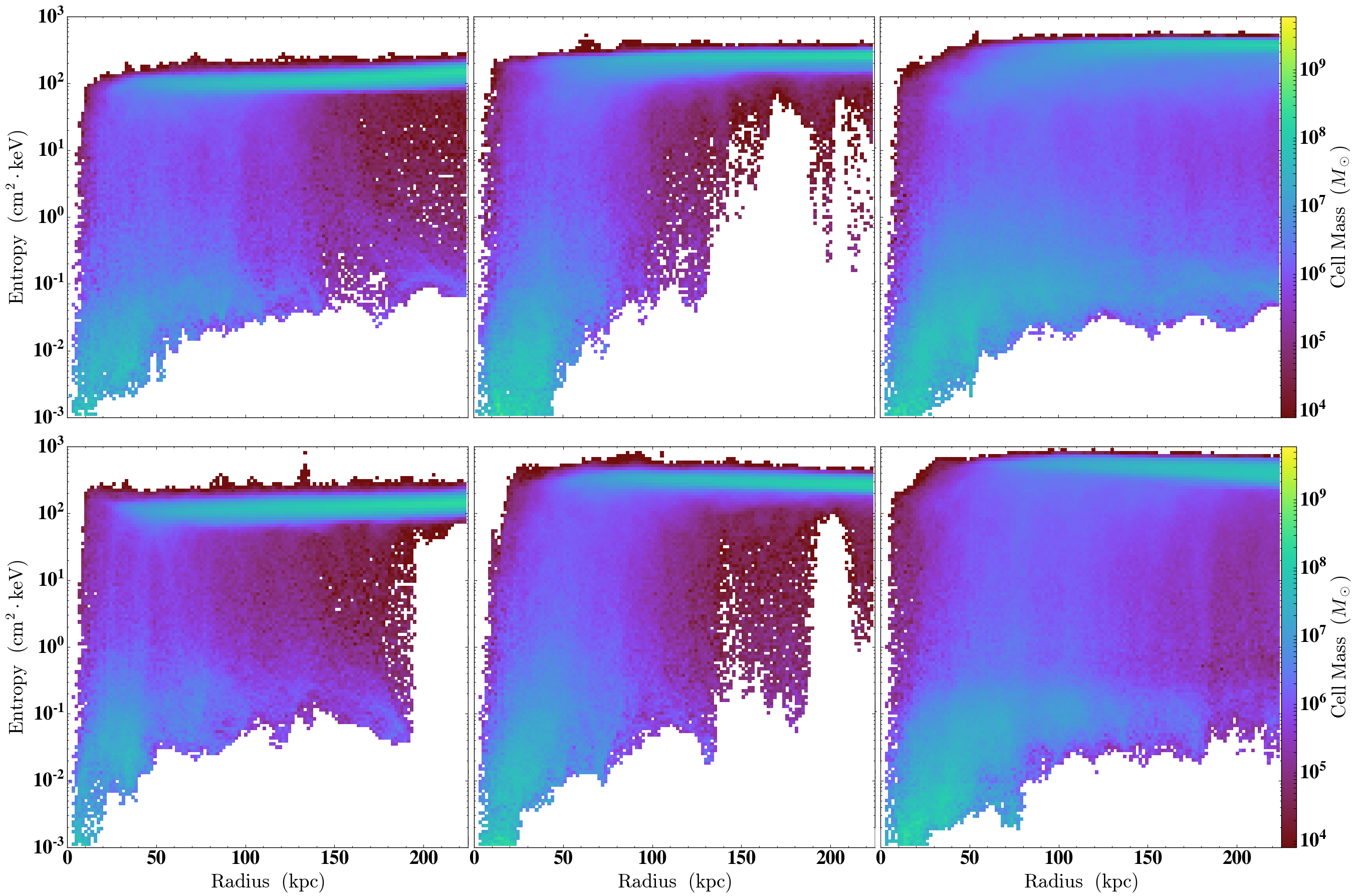}
    \caption{Entropy vs. radius showing the total mass in a bin for the Hydro Rot (top) and MHD Rot (bottom) runs. We specifically show these at 3 (left), 6 (middle), and 9 (right) Gyrs. Plots are made using 128 bins for the x and y quantities.}
    \label{fig:entropy_radial}
\end{figure*}

\end{document}